\documentclass[showpacs,showkeys,preprintnumbers,pre,amsmath,amssymb,superscriptaddress]{revtex4}
\usepackage{graphicx}
\usepackage{natbib}
\usepackage{color}
\usepackage{amsmath}
\usepackage{enumerate}
\usepackage{hyperref}
\usepackage[toc,page]{appendix}
\def\strutdepth{\dp\strutbox}
\def\nw#1{\strut\vadjust{\kern-\strutdepth\vtop to0pt{\vss\hbox to\hsize
{\hskip\hsize\hskip5pt$\leftarrow$\hss\strut}}}{\em #1}}

\usepackage[makeroom]{cancel}

\chardef\_=`_

\begin{document}

\title{A one dimensional modal approach for flows controlled by contact line motion}
\author{H. Perrin} 
\email{hugo.perrin@espci.fr}
\affiliation{Laboratoire de Physique Statistique, UMR 8550 ENS-CNRS, Univ. Paris-Diderot, 24 rue Lhomond, 75005, Paris.}

\author{D. Belardinelli} 
\email{belardinelli@roma2.infn.it}
\affiliation{$1$ Department of Physics \& INFN, University of Rome ``Tor Vergata'', Via della Ricerca Scientifica 1, 00133, Rome, Italy.}

\author{M. Sbragaglia} 
\email{sbragaglia@roma2.infn.it}
\affiliation{$1$ Department of Physics \& INFN, University of Rome ``Tor Vergata'', Via della Ricerca Scientifica 1, 00133, Rome, Italy.}

\author{B. Andreotti}
\email{andreotti@lps.ens.fr}
\affiliation{Laboratoire de Physique Statistique, UMR 8550 ENS-CNRS, Univ. Paris-Diderot, 24 rue Lhomond, 75005, Paris.}


\begin{abstract}
The hydrodynamics of a liquid-vapour interface in contact with an heterogeneous surface is largely impacted by the presence of defects at the smaller scales. Such defects introduce morphological disturbances on the contact line and ultimately determine the force exerted on the wedge of liquid in contact with the surface. From the mathematical point of view, defects introduce perturbation modes, whose space-time evolution is governed by the interfacial hydrodynamic equations of the contact line. In this paper we derive the response function of the contact line to such generic perturbations. The contact line response may be used to design simplified 1+1 dimensional models accounting for the complexity of interfacial flows coupled to nanoscale defects, yet offering a more tractable mathematical framework to include thermal fluctuations and explore thermally activated contact line motion through a disordered energy landscape.
\end{abstract}

\date{\today}%

\maketitle

\section{Introduction}
The dynamic spreading of a liquid on an heterogeneous substrate is a complex problem at the crossroads between physics, chemistry, and engineering~\cite{bonn_wetting_2009,snoeijer_moving_2013,Amberg14}. It is involved in a variety of industrial processes such as boiling enhancement~\cite{kwon2013increasing}, ink-jet printing of electronic circuits~\cite{sirringhaus2000high,van2008inkjet,ahn2009omnidirectional,onses2013hierarchical}, droplet control~\cite{bird2013reducing,liu2014pancake}, patterning of substrates~\cite{fan2000rapid,park2007high,stuart2010emerging,duprat2012wetting,galliker2012direct} and even with adhesion of reticulated polymers~\cite{paxson2013self}. The seminal problem for such dynamic contact line flows is the deposition of a thin liquid layer on a solid surface withdrawn from a liquid reservoir~\cite{E04b,E05,eggers_contact_2005,ChanPOF12}. If the hydrodynamic description of the problem in the ideal situation of a flat homogeneous solid is now well understood, the influence of roughness and of chemical defects at the nanometer scale remains a challenging problem, largely open despite recent progresses~\cite{snoeijer_moving_2013}. \\
The force balance exerted on a wedge of liquid along the contact line under the influence of the solid is macroscopically parametrized by the surface tensions of the liquid-vapor ($\gamma$), solid-liquid ($\gamma_{\mbox{\tiny sl}}$), and solid-vapor interfaces ($\gamma_{\mbox{\tiny sv}}$): as a direct consequence of the intermolecular forces, they provide excess free energies associated with the interfaces, and combine at equilibrium to provide the contact angle $\theta_Y$ made by the liquid-vapour interface with respect to the solid, i.e. the celebrated Young's law~\cite{deGe02}
\begin{equation}
\gamma \cos\theta_Y = \gamma_{\mbox{\tiny sv}}-\gamma_{\mbox{\tiny sl}}.
\end{equation}
Substrate heterogeneities are therefore naturally described in terms of a frozen, disordered surface energy landscape. This constitutes the first difficulty of the problem: the force of solid origin exerted on the fluid depends on the location of the contact line. The contact angle locally made by the liquid is selected at a molecular scale~\cite{snoeijer_microscopic_2008}, and Young's law therefore acts as a boundary condition for the mesoscopic interface.\\
The flow resulting from the contact line motion must be described by interfacial hydrodynamics, which immediately reveals the second difficulty of the problem: as a contact line is a geometrical singularity, the corner flow~\cite{huh_scriven_1971} presents a viscous stress that tends to diverge at the contact line, but remains finite due to some molecular scale regularization process. Viscous dissipation of energy takes place at all length scales between the molecular scale and the size of the meniscus~\cite{huh_scriven_1971}. This yields a total dissipation that is neither integrable at the singularity nor at infinity, and the problem requires a cut-off at both small scale and large scale. Typically, these cut-offs appear at molecular scale ($\sim 10^{-9}$m), and at the scale of the capillary length $\ell_\gamma$ ($\sim 10^{-3}$m). Each of the decades in between the microscopic scale and the macroscopic scale contributes to the viscous dissipation, revealing the intrinsic multi-scale character of wetting flows. These features of moving contact lines were first appreciated by~\cite{huh_scriven_1971}, who analytically solved the flow in a perfect wedge using similarity solutions. The equations considerably simplify in the limit of small interfacial slopes and curvatures, i.e. in the lubrication limit~\cite{ODB97}. In such limit, the relevant dynamical quantities reduce to the thickness $h$ of liquid from the solid to the interface and the average velocity ${\vec U}$ parallel to the solid (plate).\\ 
The importance of physico-chemical heterogeneities at small scales together with the necessity to include a regularization mechanism for the contact line problem set a compelling case for the understanding of the role of thermal fluctuations. Indeed, at nanoscales, the strength of thermal fluctuations becomes comparable to that of surface tension, hence fluctuations may trigger activated dynamics across defects~\cite{Prevost99}. From the point of view of macroscopic interfacial hydrodynamics, thermal fluctuations may be embedded in a continuum description of the contact line flows based on fluctuating hydrodynamics~\cite{landau_fluidmech_book,flekkoy_fluctuating_1996,zaratebook}, i.e. the equations of hydrodynamics where the viscous stress tensor is supplemented with a stochastic contribution accounting for the random motion of molecules at small scales. In the lubrication limit, some studies have been proposed in the literature~\cite{davidovitch_spreading_2005,gruen_thin-film_2006,rauscher_wetting_2008,belardinelli2016thermal}. In these studies, boundary conditions are typically needed to account for the impenetrable nature of the boundaries~\cite{nesic_dynamics_2015,nesic_nonlinear_2015}; capillary waves, in turn, may be affected by the restrictions imposed by the boundaries~\cite{fisher_walks_1984,bricmont_random_1986,lebowitz_effect_1987}, resulting in morphological changes of the average profile on the scale of the thermal length. In this framework, some phenomenological (coarse grained) parameter may be introduced~\cite{belardinelli2016thermal} to account for the affinity of the contact line with the substrate, and exact calculations may be performed to predict the shape of the profiles close to the wall. However, the microscopic derivation of such parameter requires a suitable matching with an inner description, possibly including the details of the heterogeneities. This leaves us with a third difficulty, i.e. rationalizing a framework for a fluctuating contact line problem coupled to a precise realization of nanoscale defects. \\
This paper aims to take a step further in this direction. Placing the effects of thermal fluctuations on the full (time-dependent) contact line problem with an heterogeneous realization of defects on a surface looks an hard task. This is not even required if we are not interested in all the details of  the contact line profile~\cite{Crassous}, but rather want to predict and control the activated hopping of its average position. Disturbances introduced by defects may be small, and can still be described in the hydrodynamic framework as elastic perturbations~\cite{JdG84,Crassous,deGennesRMP,deGennes_book}. Thermal activation across defects has been recently studied in a semi-phenomenological framework by~\cite{perrin_2016}, and compared to experiments. Our aim is here to provide a rigorous theoretical framework to derive further reduced models, including memory effects ignored in ~\cite{perrin_2016}. The evolution equations of such perturbations are analyzed in the lubrication limit for the well known dip-coating set-up~\cite{E04b,E05,eggers_contact_2005,ChanPOF12}, consisting of a plate withdrawal from a bath at a constant velocity. When the effects of perturbations are averaged in space along the contact line profile, we are left with a simplified 1+1 force balance equation for the time evolution of the average position coupled to the effects of heterogeneities. Crucial for our analysis is the characterization of the {\it response function}. Given a frozen energy landscape imposed by defects, the response function gives the (linear) relation between the displacement and the force exerted along the contact line profile. This force materializes in two different contributions: one is related to the deformation of the liquid interface (a memory term), and the other is the force set by the spatial variation of the contact angle. This is somehow a linear rheological characterization~\cite{Larson} of the contact line, with associated storage and loss moduli, so to say. Once the simplified model is obtained, the introduction of noise is more tractable~\cite{HanggiReview} and allows to write a Langevin equation for the flow of contact line motion coupled to defects.\\ 
The paper is organized as follows: Section \ref{sec:theoreticalframework} deals with the theoretical framework utilized in the present paper, i.e. the dip-coating geometry in the lubrication approximation; the properties of the dynamical base state in such a geometry are reviewed in Section \ref{sec:basestate} and the dynamical equations for the perturbations are the subject of Section \ref{sec:linearized}; the characterization of the response function pertains Section \ref{sec:response}, while some concluding remarks and perspectives are offered in Section \ref{sec:concluding}.

\section{Theoretical framework}\label{sec:theoreticalframework}
\subsection{Notations}
\begin{figure}[t!]
\includegraphics{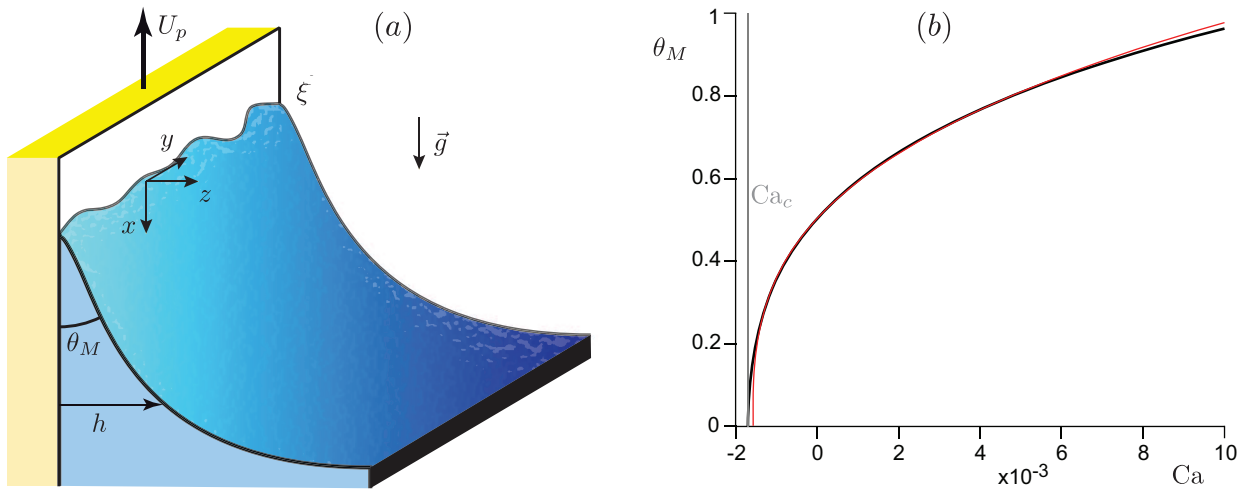}
\caption{(a) The response function of the dynamic contact line is determined in the dip-coating geometry. A vertical plate is withdrawn from a bath at the velocity $U_p$. The contact line profile is denoted by $\xi$. The bath, far from the plate, is at equilibrium. Its asymptotic profile when approaching the plate resemble that of a static bath and would join the plate, if prolonged, at an angle $\theta_M$ called the macroscopic contact angle (see Section \ref{MacroscopicContactAngle}). (b) Relation between the apparent contact angle $\theta_M$ and the capillary number ${\rm Ca} = \eta U_p/\gamma$ determined using a slip length $\ell_s=2.5\,10^{-6}\,\ell_\gamma$ and a microscopic contact angle $\theta_0=0.5\; \rm{rad}$, corresponding to about $28.6^\circ$. Here, $\rm{Ca_c}$ is the threshold capillary number below which a liquid film is entrained on the plate, and corresponds to $\theta_M=0$. The thin red line is the approximation by the Cox-Voinov formula (\ref{CV}).}
\label{Fig1}
\end{figure}

Our aim is here to describe the contact line motion (thermally activated or not) on an heterogeneous substrate. We consider the seminal dip-coating geometry in which a plate is withdrawn from a bath at a constant velocity $U_p$~\cite{E04b,E05,eggers_contact_2005,ChanPOF12} [see Fig.~\ref{Fig1}(a)]. The liquid has a mass density $\rho$, a viscosity $\eta$ and a surface tension $\gamma$. The coordinates are $x$ and $y$ along the plate, with $x$ going from the contact line to the bath, and $z$ normal to the plate, while $t$ is the instant of time. The plate velocity $U_p$ can be rescaled by the typical velocity $\gamma/\eta$ for which viscous stress and capillarity are of the same order of magnitude:
\begin{equation}
{\rm Ca} = \frac{\eta U_p}{\gamma}.
\end{equation}
Gravity, whose acceleration is $g$, fixes the outer lengthscale of the problem, the capillary length
\begin{equation}
\ell_\gamma= \sqrt{\frac{\gamma}{\rho g}}.
\end{equation}
The boundary condition is the contact angle $\theta_Y(y,t)$ along the contact line profile that we decompose as $x=\xi_0+\xi(y,t)$. $\xi_0$ is the average position over space and time (or over realizations) and $\xi$ the fluctuating part, in space and time. On the one hand, the angle of the liquid interface along the contact line $\theta_Y(y,t)$ results from a frozen landscape $T_Y(x,y)$ such that it corresponds to the value of the frozen landscape at the location of the contact line: $\theta_Y(y,t)=T_Y(\xi_0+\xi(y,t),y)$. The contact angle profile is determined by the contact line position, which itself depends on the contact angle distribution. On the other hand, the flow and therefore the evolution of the contact line position is entirely driven by $\theta_Y(y,t)$, as being the boundary condition of the dynamic liquid interface. We can therefore solve for the hydrodynamics problem, assuming that $\theta_Y(y,t)$ is known in advance and determine the evolution of $\xi(y,t)$. The two parts of the problem namely the selection of $\theta_Y(y,t)$ by the value of the frozen landscape at the location of the contact line and the hydrodynamics driven by $\theta_Y(y,t)$ can therefore be treated separately and coupled afterward.\\
We use the lubrication approximation of Navier-Stokes~\cite{ODB97} and linearize the solution with respect to the perturbation $F(y,t)=\gamma(\cos \theta_Y(y,t)-\cos \theta_0)$, where $\cos \theta_0$ is the average over space and time (or over realizations in an unsteady statistical process) of $\cos \theta_Y$, which is evaluated from the frozen field $T_Y (x, y)$ along the contact line. The value of $\cos \theta_0$ therefore results from the dynamics but the hydrodynamic problem can be treated, parametrized by $\theta_0$, ignoring its actual value. We wish to find the position, knowing the boundary condition $\theta_Y(y,t)$. For this, we perform the double Fourier transform in space and time and denote by $q$ the wavenumber and $\omega$ the angular frequency.

\subsection{Deterministic equation for the contact line}

The contact line profile $\xi(y,t)$ is decomposed over transverse modes:
\begin{equation}
\bar{\xi}(p,t)= \frac{1}{\Lambda}\int^{+\Lambda/2}_{-\Lambda/2} dy\;e^{-j2\pi py/\Lambda}\xi(y,t)
\end{equation}
where $\Lambda$ is the integral transverse lengthscale. $\bar \xi$ is homogeneous to a $\xi$ and in particular $\bar \xi(0,t)$ is the average position of the contact line over space. $\hat \xi(p,\omega)$ denotes the Fourier transform in space and time:
\begin{equation}
\hat{\xi}(p,\omega)= \frac{1}{\Lambda}\int^{+\infty}_{-\infty} dt\int^{+\Lambda/2}_{-\Lambda/2} dy\;e^{-j\omega t-j2\pi py/\Lambda}\xi(y,t)
\end{equation}
or, equivalently:
\begin{equation}
\xi(y,t)=\sum_{p=-\infty}^{\infty} \int^{+\infty}_{-\infty} \frac{d\omega}{2\pi}\;e^{j\omega t+j2\pi p y/\Lambda}\hat{\xi}(p,\omega).
\end{equation}
It is linearly related to the force disturbance Fourier transform:
\begin{equation}
\hat F(p,\omega)=\gamma {\mathcal C}(2\pi p/\Lambda,\omega)\hat \xi(p,\omega)
\label{psieq}
\end{equation}
where the {\it response function} ${\mathcal C}(q,\omega)$, which is homogeneous to the inverse of a length, is calculated from hydrodynamics for any mode
\begin{equation}
q=\frac{2\pi}{\Lambda} p.
\end{equation}
The average position of the contact line $\xi_0$ over space and time has been singled out, which is actually a function of $\theta_0$ and of the capillary number ${\rm Ca}$. Note that the problem does not need to be in a statistically steady state: $\theta_0$ and ${\rm Ca}$ may vary as a function of time, in which case $\xi_0$ adapts consequently (more details in Section~\ref{sec:dependenceonCa}). Further note that $\theta_0$ is ultimately a function of ${\rm Ca}$ and of the statistical properties of the frozen disorder $T_Y$, due the fact that it results from an average along the contact line. The sampling of the energy landscape is therefore biased. Similarly, Eq.~\eqref{psieq} is the dynamical equation for the contact line fluctuations, as $F$ is a function of $\xi$ through the reading of the energy landscape. We then rewrite the response function ${\mathcal C}$ using the following decomposition between real (storage modulus) and imaginary (loss modulus) part:
\begin{equation}\label{eq:Cdecomposition}
{\mathcal C}(q,\omega)= \frac{1}{\ell_\gamma\hat \phi_q(\omega)}\left[\hat \psi_q(\omega)+j\frac{\eta \ell_\gamma}{\gamma} \omega\right]
\end{equation}
where the capillary length $\ell_\gamma$ is chosen as a typical lengthscale and $\gamma/\eta \ell_\gamma$ as a typical frequency. Then, both $\hat \psi_q$ and $\hat \phi_q$ are dimensionless functions. The angular frequency $\omega$, rescaled by $\gamma/\eta \ell_\gamma$, will be noted, in the rest of the article,
\begin{equation}
\Omega \equiv \frac{\eta \ell_\gamma}{\gamma} \omega
\end{equation}
so that $\ell_\gamma {\mathcal C}= (\hat \psi_q+j\Omega)/\hat \phi_q$. The equation reads, in the real time for every $q$ mode,
\begin{equation}\label{eq:deterministic_eq}
\eta \frac{d\bar \xi}{dt} + \frac{\gamma}{\ell_\gamma}\int_{-\infty}^{t}  \psi_q(t-t') \bar \xi(t') dt'= \gamma \int_{-\infty}^{t} \phi_q(t-t') \left(\overline{\cos \theta_Y}-\cos \theta_0\right) dt'.
\end{equation}
%
The first term is, as requested, the evolution operator, which result from the dissipation induced by the contact line motion. It is naturally proportional to viscosity. The second term, on the left hand side is a restoring force associated with the deformations of the liquid interface induced by the contact line. It is a memory term which depends on the past deformations. As the system is causal --~only past determines future~-- $\hat \psi_q(\omega)$ and $\hat \phi_q(\omega)$ are not independent and must obey a Kramers-Kronig relation (see~\cite{Toll} and references therein). This restoring force is naturally proportional to $\gamma/\ell_\gamma$. The term on the right hand side is the forcing by the solid heterogeneities: $\overline{\cos \theta_Y}$ (times $\gamma$) denotes the Fourier transform in space of the force per unit line exerted under the influence of the solid: 
\begin{equation}
\overline{\cos \theta_Y}(p,t)= \frac{1}{\Lambda}\int^{+\Lambda/2}_{-\Lambda/2} dy\;e^{-j2\pi py/\Lambda} \cos (T_Y(\xi_0+\xi(y,t),y)).
\end{equation}
The time kernel $ \phi_q$ originates from the viscous damping of perturbation, which leads to a time-memory of the energy landscape seen by the contact-line. Finally, notice that for a small variation  $\theta_1$ of the contact angle to its space and time average one can linearize the force disturbance as $\cos (\theta_0+\theta_1) -\cos \theta_0 \simeq - \sin \theta_0\,\theta_1$, hence from \eqref{psieq} we can write
\begin{equation}\label{eq:contactvsFORCE}
{\mathcal C}(2\pi p/\Lambda,\omega)\hat \xi(p,\omega) \simeq -\sin \theta_0 \, \hat{\theta}_1(p,\omega)
\end{equation}
which relates the contact angle variation and the contact line disturbance through the response function.

\subsection{Stochastic equation, with thermal noise}
We wish now to determine the thermal noise that can be added to the dynamical equation, applying the fluctuation dissipation theorem~\cite{HanggiReview}. The equation reads in the real time:
\begin{equation}\label{eq:Langevin}
\eta \frac{d\bar \xi}{dt} + \frac{\gamma}{\ell_\gamma}\int_{-\infty}^{t}  \psi_q(t-t') \bar \xi(t') dt'= \gamma \int_{-\infty}^{t} \phi_q(t-t') \left(\overline{\cos \theta_Y}-\cos \theta_0\right) dt'+\sqrt{\frac{\eta k_B T}{\ell_\gamma}}\,W(t)
\end{equation}
where $W$ is some zero-mean random Gaussian forcing to be prescribed and depends on the $q$ mode. This noise is rescaled by the product of the Boltzmann constant $k_B$ by the temperature $T$, by the viscosity $\eta$ and by the characteristic length $\ell_\gamma$. Provided the knowledge of the equilibrium correlations $\left<|\bar \xi|^2\right>$, one finds in the absence of forcing (see~\cite{Zwanzig} and references therein)
\begin{equation}
\left<W(t)W(t')\right> = \frac{\eta \ell_\gamma}{k_B T} \left<|\bar \xi|^2\right> \psi_q(|t-t'|).
\end{equation}
In first approximation, we can relate $\left<|\bar \xi|^2\right>$ to  the fluctuation $\delta h$ of the height $h(x,y)$ from the plate. The structure of guided modes on a static meniscus can in principle~\cite{belardinelli2016thermal} be calculated but we wish here to get a simple first approximation, correct from the scaling point of view and the order of magnitude. We therefore hypothesize that the meniscus behaves like a flat interface bounded by the bath and by the contact line, assumed to behave like two rigid boundaries. We therefore introduce the length $L_\gamma$ along the meniscus. By its very definition, the distance of the (average) contact line above the bath is related to the macroscopic angle $\theta_M$ by $\sqrt{2(1-\sin \theta_M)} \ell_\gamma$ (see Section \ref{MacroscopicContactAngle}). Approximating the meniscus by a plane, we therefore get:
\begin{equation}
L_\gamma = \frac{\sqrt{2(1-\sin \theta_M)}}{\cos \theta_M} \ell_\gamma.
\end{equation}
By double Fourier transforming $\delta h(x,y)$ we get $\delta\bar h(n,p)$, and from equipartition of energy~\cite{Saf03} we can write 
\begin{equation}
\left<|\delta\bar h|^2\right> = \frac{k_BT}{\gamma L_\gamma\Lambda(k^2+q^2)}.
\end{equation}
Here, $h(x,y)$ is treated as a free surface, $k=2\pi n/L_\gamma$ and $q=2\pi p/\Lambda$ are the wavenumbers respectively in the direction of gravity and in the transversal direction (along the contact line). Note that the mode with both $p=0$ and $n=0$ equals the spatial average of $\delta h$, so that it identically vanishes by mass conservation. Then, if $q\neq0$, by summing over all the integers $n$ we have, within a proportionality coefficient $f$ depending on the characteristics of molecular interactions close to the plate~\cite{belardinelli2016thermal}:
\begin{equation}
\left<|\bar \xi|^2\right> = \frac{f}{\tan^2\theta_0} \sum_{n=-\infty}^{\infty} \left<|\delta\bar h|^2\right> = \frac{fk_BT}{4\pi\gamma\tan^2\theta_0} \frac{\coth(\pi pL_\gamma/\Lambda)}{p}
\end{equation}
so that
\begin{equation}\label{eq:noise_form}
\left<W(t)W(t')\right> = \frac{f\eta \ell_\gamma}{2\gamma\tan^2\theta_0} \frac{\coth(qL_\gamma/2)}{q\Lambda} \psi_q(|t-t'|).
\end{equation}
In the limit $L_\gamma/\Lambda\gg1$, the expression simplifies into:
\begin{equation}
\left<W(t)W(t')\right> = \frac{f\eta \ell_\gamma}{2\gamma |q|\Lambda\tan^2\theta_0} \psi_q(|t-t'|).
\end{equation}
Instead, for the $q=0$ mode we have to exclude $n=0$ from the counting, to get
\begin{equation}
\left<|\bar \xi|^2\right> = \frac{2f}{\tan^2\theta_0} \sum_{n=1}^{\infty} \left<|\delta\bar h|^2\right> = \frac{fk_BTL_\gamma}{12\gamma\Lambda\tan^2\theta_0}
\end{equation}
and hence
\begin{equation}\label{eq:noise_formq=0}
\left<W(t)W(t')\right> = \frac{f\eta \ell_\gamma L_\gamma}{12\gamma\Lambda\tan^2\theta_0} \psi_{0}(|t-t'|).
\end{equation}
%

\section{Dynamical base state}\label{sec:basestate}
\subsection{Lubrication equations}

The lubrication equations with a Navier slip boundary condition read~\cite{ODB97}:
\begin{eqnarray}
\partial_t h+ {\vec \nabla} \cdot ( h \, {\vec U}) &=& 0,\label{massdim}\\
\gamma {\vec \nabla} \kappa  +\rho g \vec e_x+ \frac{3\eta (U_p{\vec e_x}-{\vec U})}{h(h+3\ell_s)} &=& {\vec 0}\label{momentumdim}
\end{eqnarray}
where $\ell_s$ is the slip length. They constitute a controlled approximation of Stokes equations under the condition of small slope and small product of the curvature $\kappa$ by the thickness $h$.
Here the surface is located at $z=h(x,y,t)$, while ${\vec \nabla}=\vec e_x\partial_x + \vec e_y\partial_y$ is the gradient operator along the plate, $\vec e_i$ being the unit vector of the $i$-th coordinate, ${\vec U}(x,y,t)=\vec e_x U_x(x,y,t) + \vec e_y U_y(x,y,t)$ is the velocity along the plate averaged over $z$, that is
\begin{equation}
{\vec U} = \frac{1}{h}\int_0^h {\vec u} \, dz
\end{equation}
${\vec u}(x,y,z,t)$ being the true hydrodynamic velocity along the plate. The continuity equation \eqref{massdim} expresses the conservation of mass for the problem at hand, and the quantity $h\,{\vec U}$ is the flux vector along the plate. The curvature $\kappa(x,y,t)$ of the surface, appearing in the force balance Eq.~\eqref{momentumdim}, is given by
\begin{equation}
\label{curvature}
\kappa = \frac{\left(1+ \partial_{y} h^2\right)\partial_{xx}h +\left(1+ \partial_{x} h^2\right)\partial_{yy}h-2\partial_x h \, \partial_y h \, \partial_{xy}h}{(1 + \partial_x h^2 + \partial_y h^2)^{3/2}}
\end{equation}
and is related to the pressure $P(x,y,t)$ by the Laplace formula $P=-\gamma \kappa$. We introduce $\gamma/\eta$ as a unit velocity and $\ell_\gamma$ as a unit length for spatial coordinates $x$ and $y$ and for the thickness $h$. Dimensionless variables will be noted in the same way as the variables themselves, except $U_p$ which becomes ${\rm Ca}$. When needed, we will give back expressions with their dimensions, mentioning it explicitly. The lubrication equations read: 
\begin{eqnarray}
\label{lubmass}
\partial_t h + {\vec \nabla} \cdot ( h \, {\vec U}) &=& 0,\\
\label{lubmom}
{\vec \nabla} \kappa  + \vec e_x+ \frac{3 ({\rm Ca}\,{\vec e_x}-{\vec U})}{h(h+3\ell_s)} &=& \vec 0.
\end{eqnarray}
The formulation of the boundary conditions for the dynamical problem requires a further discussion, and we postpone it to Section \ref{sec:BC}. Let $h(x,y,t)=h_0(x)$ be the steady transversely invariant surface profile, for which the curvature $\kappa(x,y,t)=\kappa_0(x)$ reduces to (prime $'$ means derivation with respect to $x$)
\begin{equation} 
\kappa_0=\frac{h_0''}{(1 + h_0'^2)^{3/2}}.
\end{equation}
This steady and transversely invariant liquid interface profile constitute the base state, noted with the subscript 0. From the continuity equation and condition of zero flux at the contact line we get ${\vec U}(x,y,t)={\vec U}_0(x)=\vec 0$. The lubrication equations reduce to:
\begin{equation}
\kappa'_0 + 1 + \frac{3 {\rm Ca}}{h_0(h_0 + 3\ell_s)} = 0.
\end{equation}
Regardless the convention for $\xi_0$, it is convenient to choose the location of the contact line for the steady transversely invariant case as $x=0$. The boundary conditions at the plate are then:
\begin{eqnarray}
h_0(0)&=&0, \\
h'_0(0) &=& \tan \theta_0.
\end{eqnarray}
%

\subsection{Asymptotics of the base state at the plate}

By starting from the boundary conditions for $h_0$ and $h_0'$ at $x=0$, one can find the asymptotics of the base state at the plate by solving recursively the system
\begin{eqnarray}
h_0'' &=& (1 + h_0'^2)^{3/2} \kappa_0,\\
\kappa'_0 &=& - 1 - \frac{3 {\rm Ca}}{h_0(h_0 + 3\ell_s)}.
\end{eqnarray}
We get the following asymptotics of the base state at the plate, for $x\to0$,
\begin{eqnarray}
\label{eq:baseh}h_0(x)&=& t_0 x - \frac{{\rm Ca}\,(1 + t_0^2)^{3/2}}{2\ell_s t_0} x^2\ln\left(\frac{x}{\ell}\right)+\frac{3{\rm Ca}\,(1 + t_0^2)^{3/2}}{4\ell_s t_0} x^2 + O(x^3\ln^2x)\\
\label{eq:baseh'}h_0'(x)&=& t_0 - \frac{{\rm Ca}\,(1 + t_0^2)^{3/2}}{\ell_s t_0}x\ln\left(\frac{x}{\ell}\right)+\frac{{\rm Ca}\,(1 + t_0^2)^{3/2}}{\ell_s t_0}x + O(x^2\ln^2x)\\
\label{eq:basek}\kappa_0(x)&=& - \frac{{\rm Ca}}{\ell_s t_0}\ln\left(\frac{x}{\ell}\right) - \frac{{\rm Ca}^2(1 + t_0^2)^{3/2}}{2\ell_s^2 t_0^3}x\ln\left(\frac{x}{\ell}\right) -\left[1 - \frac{{\rm Ca}}{3\ell_s^2} - \frac{5{\rm Ca}^2(1 + t_0^2)^{3/2}}{4\ell_s^2 t_0^3}\right]x + O(x^2\ln^2x)
\end{eqnarray}
where $t_0=\tan\theta_0$, for shortness, while $\ell$ is a free parameter, adjusted by shooting to match the bath. Note that, in practice, it is convenient to introduce the quantity ${\rm Ca} \ln \ell$ rather than $\ell$ itself.

\subsection{Macroscopic contact angle}
\label{MacroscopicContactAngle}
The concept of macroscopic contact angle has long been a source of confusion in the literature. The proper way of defining it is to start from the asymptotics at the bath, which is exactly that of a static bath at equilibrium. The constant $M$ (see below), though, which describes the exponential departure from the flat bath (considering $x$ vs $h_0$) depends on the dynamic solution between the scale of the slip length and that of the capillary length. The static-like asymptotics at the bath can be prolonged and would join the plate at an angle, which is by definition the macroscopic angle $\theta_M$. The macroscopic contact angle is therefore not a true interface angle but is defined by asymptotic matching of the solution coming from the dynamical range of scales with an outer, static bath solution. The position at which the static bath interface would join the plate is almost the same as the true contact line position. In practice, the macroscopic angle $\theta_M$ can therefore be defined from the altitude $\delta$ of the contact line using the following static relation~\cite{landau_fluidmech_book}: 
\begin{eqnarray}\label{eq:xcl}
\delta=\sqrt{2(1-\sin \theta_M)}\,\ell_\gamma.
\end{eqnarray}
The difference between the capillary forces $\gamma \cos\theta_M$ and $\gamma \cos\theta_0$ at macroscopic and microscopic scales results from the viscous force integrated along the plate, between the inner and outer scales. A useful approximation is that provided by the Cox-Voinov derivation~\cite{COXVOINOV}, which is a particular solution of the lubrication equations matched macroscopically to a vanishing curvature interface. It is derived at the linear order in angle but turns out to provide an excellent non-linear fit of the actual solution [see Fig.~\ref{Fig1}(b)], for uncontrolled reasons:
\begin{equation}
\theta_M^3 \sim \theta_0^3+9 {\rm Ca} \ln{\left(\frac{\alpha \ell_\gamma}{3\ell_s}\right)}
\label{CV}
\end{equation}
where $\alpha \simeq 0.02$ is independent, in first approximation, of the contact angle. The interested reader may find the correct asymptotic expansion in~\cite{E04b,E05,eggers_contact_2005,ChanPOF12}. Equation \eqref{eq:xcl} is an exact result for the static bath, as the static profile decays exponentially as $\delta-x\sim e^{-h}$ at large heights $h$. We therefore look for the following asymptotics at the bath for $x\to \delta$:
\begin{eqnarray}
\label{eq:bathh}h_0(x)&\sim& - \ln\left(\frac{\delta-x}{M}\right)\\
\label{eq:bathh'}h_0'(x)&\sim& \frac{1}{\delta-x}\\
\label{eq:bathk}\kappa_0(x)&\sim& \delta-x
\end{eqnarray}
where $M$ is a free parameter, adjusted by shooting to match the contact line. Note that $\delta$ is the distance between the average contact line and the bath.

\section{Linearized equations}\label{sec:linearized}
\subsection{Governing equations}
We linearize Eqs.~\eqref{curvature}-\eqref{lubmom} about the basic profile $h_0(x)$, writing the profile as a combination of the base state and a small disturbance, noted with subscript 1, oscillating in time at frequency $\Omega$ and modulated spatially at wave number $q$
\begin{eqnarray}
\label{perturbationfull}
h(x,y,t)&=&h_0(x)+ \, h_1(x)\,e^{j \Omega t + jqy} \\ 
\kappa(x,y,t) &=&\kappa_0(x)+  \, \kappa_1(x) \,e^{j \Omega t + jqy} \\
U_x(x,y,t) &=&  \, u_1(x) \,e^{j \Omega t + jqy}\\
U_y(x,y,t) &=&  \, v_1(x) \,e^{j \Omega t + jqy}.
\end{eqnarray}
The curvature linearizes into:
\begin{equation}
\kappa_1 = -\frac{ q^2h_1}{(1+{h_0'}^2)^{1/2}} + \frac{ h_1'' }{(1+{h_0'}^2)^{3/2}} -\frac{3 \kappa_0 h_0'h_1'}{1 + {h_0'}^2}.
\end{equation}
From the $y$-component of Eq.~\eqref{lubmom}, one can eliminate $v_1$ in terms of $\kappa_1$, as 
\begin{equation}
v_1 =\,\frac{1}{3} jq h_0\left( h_0 + 3\ell_s \right) \,\kappa_1.
\end{equation}
It is convenient to introduce the variable 
\begin{equation}
\mathcal{F}_1(x) =h_0(x) u_1(x)
\end{equation}
which represents the flux in the $x$ direction at first order (the zeroth order flux being zero). We get, from the linearized lubrication equations about the base state, the differential equations obeyed by the disturbed liquid interface~\cite{snoeijer_2007b}:
\begin{eqnarray}\label{eq:linearized1}
h_1'' &=& (1+{h_0'}^2)q^2h_1+3 (1 + {h_0'}^2)^{1/2} \kappa_0 h_0'h_1'+ (1+{h_0'}^2)^{3/2} \kappa_1\\
\kappa_1' &=& \frac{3 {\rm Ca}(2h_0+3\ell_s)}{h_0^2 (h_0+3\ell_s)^2} h_1+ \frac{3 }{h_0^2(h_0 +3 \ell_s)} \mathcal{F}_1 \label{eq:linearized2} \\
\mathcal{F}_1' &=&-j \Omega h_1+ \frac{h_0^2\left( h_0 + 3\ell_s \right)q^2}{3}\kappa_1. \label{eq:linearized3}
\end{eqnarray}
Defining the quadrivector ${\cal X}$ as
\begin{equation}
{\cal X} =  \left(\begin{array}{c}  h_1\\ h_1'  \\ \kappa_1\\\mathcal{F}_1\end{array} \right) 
\end{equation}
one can rewrite the linearized equations \eqref{eq:linearized1}-\eqref{eq:linearized3} as~\cite{snoeijer_2007b}
\begin{equation}
\label{linearized}
\frac{d{\cal X}}{dx} = {\cal M}{\cal X}
\end{equation}
where
\begin{eqnarray}
{\cal M} =\left(\begin{array}{cccc} 0 & 1 & 0 & 0 \\ 
&&&\\
(1+{h_0'}^2)q^2 & 3 (1 + {h_0'}^2)^{1/2} \kappa_0 h_0' & (1+{h_0'}^2)^{3/2} & 0 \\
&&&\\
\frac{3 {\rm Ca}(2h_0+3\ell_s)}{h_0^2 (h_0+3\ell_s)^2}& 0 & 0 & \frac{3 }{h_0^2(h_0 +3 \ell_s)} \\
&&&\\
-j \Omega & 0 &  \frac{h_0^2\left( h_0 + 3\ell_s \right)q^2}{3}   & 0 \end{array}\right). \nonumber 
\end{eqnarray}

\subsection{Boundary conditions and asymptotics at the plate}\label{sec:BC}
There are many formulations that are equivalent at the linear order but which are differently accurate at the non-linear order. In particular the exact solution of the problem exactly reduce to the solution of the linear problem at the linear order. Let us introduce the following formulation:
\begin{equation}
h(x,y,t)=h_0(x-\xi)+ \tilde h_1(x-\xi)e^{j \Omega t + jqy}
\end{equation}
where (the real part of) $\xi=\hat{\xi}e^{j \Omega t + jqy}$ parametrizes the disturbance to the contact line position for given frequency $\Omega$ and wavenumber $q$. The boundary conditions are:
\begin{eqnarray}\label{bound0}
h(\xi,y,t)&=&0\\
\partial_x h(\xi,y,t)&=& \tan \theta_Y(y,t)\simeq \tan \theta_0+\left. \frac{d \tan \theta}{d \theta} \right|_{\theta=\theta_0}\theta_1(y,t) \\
\mathcal{F}(\xi,y,t)&=&0.
\end{eqnarray}
Note that the contact angle is normally taken along the normal to the contact line. However, as the base state is invariant along the $y$ direction, the normal is along $x$ within negligible quadratic disturbances, hence the above expressions. We see that the description in terms of the displaced variable $x-\xi$ is perfectly well behaved. The linear equations giving $\tilde h_1$ are entirely equivalent to those giving $h_1$. $\tilde h_1$ is totally equivalent to $h_1$, at the linear order (but not at the non-linear order). We shall therefore use $h_1$, which leads to simpler equations but keep in mind that we will actually represent the solution by $\tilde h_1$. The equivalence is given by the following equations:
\begin{equation}
h_0(x)+ h_1(x)e^{j \Omega t + jqy} = h(x,t)\simeq  h_0(x)-\xi h'_0(x)+\tilde h_1(x)e^{j \Omega t + jqy}
\end{equation}
from which we get:
\begin{equation}
\tilde h_1(x)=h_1(x)+\hat \xi h'_0(x).
\end{equation}
Therefore, based on \eqref{eq:contactvsFORCE}, we can write:
\begin{eqnarray}
h_1(0)&=& - \hat \xi \tan{\theta_0}\\
\label{eq:BCh1'}\lim_{x \to 0} (h_1'(x)+\hat \xi h''_0(x))&=& - \hat \xi\frac{1 + \tan^2 {\theta_0}}{\sin {\theta_0}} {\mathcal C} \\
\mathcal{F}_1(0)&=&0.
\end{eqnarray}
Note that $h''_0(x)$ diverges logarithmically as $h_0''(x)\sim  - \frac{{\rm Ca}(1+\tan^2\theta_0)^{3/2}}{\ell_s \tan \theta_0} \ln\left(\frac{x}{\ell}\right)$ at $x=0$, which explains the formulation of \eqref{eq:BCh1'} with the limit. Here ${\mathcal C}$ is the response function, expressing the ration between the disturbance of the forcing $\cos\theta_Y-\cos\theta_0$ and the disturbance on contact line position. As it is linear, the solution is independent of the amplitude of the disturbance $\hat \xi$ so that we can take a unit $\hat \xi$ without loss of generality.\\
We now wish to derive the general asymptotics at the plate, to determine which of them are consistent with the boundary conditions in order to perform a numerical integration. We use for the base state the approximations \eqref{eq:baseh}-\eqref{eq:basek}. We then find the asymptotics of the perturbations for $x\to0$ by solving recursively the resulting approximated system \eqref{linearized}. Let us introduce the following shorthand notations:
\begin{equation}\label{eq:short}
t_0=\tan\theta_0, \quad s_0=\sin\theta_0, \quad c_0=\cos\theta_0=(1+t_0^2)^{-1/2}.
\end{equation}
By keeping vanishing the zeroth order (in $x$) of the asymptotic solution ${\cal X}$ except for its $h_1$ component, we get
\begin{equation}
\begin{split}{\cal X}_h &= \left(\begin{array}{c}1-\tfrac{{\rm Ca}}{\ell_s s_0^2c_0}x\ln\left(\tfrac{x}{\ell}\right)+\tfrac{{\rm Ca}}{\ell_s s_0^2c_0}x+\tfrac{3{\rm Ca}^2}{2\ell_s^2 s_0^2c_0^2}x^2\ln^2\left(\tfrac{x}{\ell}\right)-\left(3+\tfrac{1}{4s_0^2}\right)\tfrac{{\rm Ca}^2}{\ell_s^2 s_0^2c_0^2}x^2\ln\left(\tfrac{x}{\ell}\right)\\
+ \left[\tfrac{q^2}{2c_0^2}+\left(1+\tfrac{1}{4s_0^2}\right)\tfrac{3{\rm Ca}^2}{2\ell_s^2 s_0^2c_0^2}\right]x^2+O(x^3\ln^3x)\\ 
\\
-\tfrac{{\rm Ca}}{\ell_s s_0^2c_0}\ln\left(\tfrac{x}{\ell}\right) +\tfrac{3{\rm Ca}^2}{\ell_s^2 s_0^2c_0^2}x\ln^2\left(\tfrac{x}{\ell}\right)- \left(3+\tfrac{1}{2s_0^2}\right)\tfrac{{\rm Ca}^2}{\ell_s^2 s_0^2c_0^2}x\ln\left(\tfrac{x}{\ell}\right) + \left(\tfrac{q^2}{c_0^2}+\tfrac{{\rm Ca}^2}{2\ell_s^2 s_0^4c_0^2}\right)x+O(x^2\ln^3x)\\
\\
-\tfrac{{\rm Ca}}{\ell_s t_0^2}x^{-1}-\tfrac{{\rm Ca}^2\,c_0}{2\ell_s^2 s_0^4}\ln\left(\tfrac{x}{\ell}\right) +O(x\ln^2x)\\
\\
-\tfrac{q^2{\rm Ca}}{2}x^2 +\left(\tfrac{q^2{\rm Ca}^2}{6\ell_s s_0^2c_0}-\tfrac{\Omega^2}{6\ell_s s_0^2c_0}\right)x^3\ln\left(\tfrac{x}{\ell}\right)+\left(\tfrac{11\Omega^2}{36\ell_s s_0^2c_0} -\tfrac{q^2{\rm Ca}\,s_0}{9\ell_s c_0^4}-\tfrac{5q^2{\rm Ca}^2}{9\ell_s s_0^2c_0}\right) x^3 +O(x^4\ln^2x)\end{array}\right)\\
&\quad- j \Omega\left(\begin{array}{c} \tfrac{1}{2\ell_s s_0^2c_0}x^2\ln\left(\tfrac{x}{\ell}\right)-\tfrac{3}{4\ell_s s_0^2c_0} x^2+O(x^3\ln^2x)\\ 
\\
\tfrac{1}{\ell_s s_0^2c_0}x\ln\left(\tfrac{x}{\ell}\right)-\tfrac{1}{\ell_s s_0^2c_0}x+O(x^2\ln^2x)\\
\\
\tfrac{1}{\ell_s t_0^2}\ln\left(\tfrac{x}{\ell}\right) +O(x\ln x)\\
\\
x-\tfrac{{\rm Ca}}{2\ell_s s_0^2c_0}x^2\ln\left(\tfrac{x}{\ell}\right) +\tfrac{3{\rm Ca}}{4\ell_s s_0^2c_0}x^2+\tfrac{{\rm Ca}^2}{2\ell_s^2 s_0^2c_0^2}x^3\ln^2\left(\tfrac{x}{\ell}\right)+\left[\tfrac{q^2}{3} -\left(4+\tfrac{1}{4s_0^2}\right)\tfrac{{\rm Ca}^2}{3\ell_s^2 s_0^2c_0^2}\right]x^3\ln\left(\tfrac{x}{\ell}\right)\\
+\left[\left(\tfrac{3}{2c_0^2}-1\right)\tfrac{q^2}{9}+\left(17+\tfrac{11}{4s_0^2}\right)\tfrac{{\rm Ca}^2}{18\ell_s^2 s_0^2c_0^2}\right]x^3+O(x^4\ln^3x)\end{array}\right).\end{split}
\end{equation}
By keeping vanishing the zeroth order of ${\cal X}$ except for its $h_1'$ component, we get
\begin{equation}
\begin{split}{\cal X}_\theta &= \left(\begin{array}{c} x+\left(\tfrac{1}{t_0^2}-2\right)\tfrac{{\rm Ca}}{2\ell_s c_0}x^2\ln\left(\tfrac{x}{\ell}\right)+ \left(2-\tfrac{1}{t_0^2}\right)\tfrac{3{\rm Ca}}{4\ell_s c_0}x^2+O(x^3\ln^2x)\\
\\
1+\left(\tfrac{1}{t_0^2}-2\right)\tfrac{{\rm Ca}}{\ell_s c_0}x\ln\left(\tfrac{x}{\ell}\right)+ \left(2-\tfrac{1}{t_0^2}\right)\tfrac{{\rm Ca}}{\ell_s c_0}x+O(x^2\ln^2x)\\
\\
\tfrac{{\rm Ca}}{\ell_s t_0^2}\ln\left(\tfrac{x}{\ell}\right) +O(x\ln^2x)\\
\\
\tfrac{q^2{\rm Ca}}{3}x^3\ln\left(\tfrac{x}{\ell}\right)-\tfrac{q^2{\rm Ca}}{9}x^3 +O(x^4\ln^2x)\end{array}\right)\\
&\quad- j \Omega\left(\begin{array}{c}O(x^3\ln^2x)\\ 
\\
O(x^2\ln^2x)\\
\\
O(x\ln x)\\
\\
\tfrac{1}{2}x^2+\left(\tfrac{1}{t_0^2}-2\right)\tfrac{{\rm Ca}}{6\ell_s c_0}x^3\ln\left(\tfrac{x}{\ell}\right)+\left(2-\tfrac{1}{t_0^2}\right)\tfrac{11{\rm Ca}}{36\ell_s c_0} x^3+O(x^4\ln^2x)\end{array}\right).\end{split}
\end{equation}
By keeping vanishing the zeroth order of ${\cal X}$ except for its $\kappa_1$ component, we get
\begin{equation}
{\cal X}_\kappa = \left(\begin{array}{c} \tfrac{1}{2c_0^3}x^2+O(x^3\ln^2x)\\ 
\\
\tfrac{1}{c_0^3}x+O(x^2\ln^2x)\\
\\
1+O(x\ln^2x)\\
\\
\tfrac{q^2\ell_s t_0^2}{3}x^3+O(x^4\ln^2x)\end{array}\right)- j\Omega \left(\begin{array}{c}o(x^3\ln^2 x)\\ 
\\
o(x^2\ln^2 x)\\
\\
o(x\ln x)\\
\\
\tfrac{1}{6c_0^3}x^3+O(x^4\ln^2x)\end{array}\right).
\end{equation}
We do not report the analogous asymptotic having the zeroth order of the $\mathcal{F}_1$ component of ${\cal X}$ non-vanishing being it unphysical (the flux must vanish at the contact line). So we must start the numerical integration with 
\begin{equation}\label{asymplate}
{\cal X}= - \tan{\theta_0} {\cal X}_h  - \frac{(1 + \tan^2 {\theta_0})^{3/2}}{\tan {\theta_0}} {\mathcal C}  {\cal X}_\theta + K {\cal X}_\kappa
\end{equation}
which satisfies the boundary conditions at the plate. The response function ${\mathcal C}$ appears in the asymptotics \eqref{asymplate}, together with the another constant $K$. Both ${\mathcal C} $ and $K$ must be chosen with a proper matching with the asymptotics at the bath. The latter will be detailed in the next subsection.

\subsection{Asymptotics at the bath}
The system \eqref{linearized} with the base state approximated by \eqref{eq:bathh}-\eqref{eq:bathk} reads :
\begin{eqnarray}\label{asymptbath}
\mathcal{F}_1' &=&-j \Omega h_1 -\frac{q^2 }{3 \ln^3(\frac{\delta-x}{M})} \,\kappa_1 \\
\kappa_1' &=&- \frac{6 {\rm Ca}}{\ln^3(\frac{\delta-x}{M})} h_1 -\frac{3}{\ln^3(\frac{\delta-x}{M})}\mathcal{F}_1 \\
h_1'' &=& \frac{q^2}{(\delta-x)^2}h_1+\frac{3 }{\delta-x}h_1'+\frac{1}{(\delta-x)^3} \kappa_1.
\end{eqnarray}
These equations admit four asymptotics, amongst which two lead to a divergence of $h_1$ as $(\delta-x)^{-1}$ and $(\delta-x)^{-2}$, respectively. The two admissible asymptotics lead to a finite value of $h_1$ (convergence as $1/\ln^2((\delta-x)/M)$ to zero and to a non-vanishing constant). As a simple criterion, we retain that $(\delta-x) h_1'$ tends to $0$ for the two admissible solutions, as $1/\ln^3((\delta-x)/M)$, but diverges for the two asymptotics that must be rejected, as $(\delta-x)^{-1}$ and $(\delta-x)^{-2}$ respectively. Furthermore, the curvature $\kappa_1$ of the two admissible solutions tend to $0$ as $(\delta-x)/\ln^3((\delta-x)/M)$ while, for one asymptotics that must be rejected tends to a constant and the other diverge as $(\delta-x)^{-1}/\ln^3((\delta-x)/M)$. The specific feature of the acceptable asymptotics is the vanishing value of both $(\delta-x) h_1'$ and $\kappa_1$. We therefore use this property in the numerics to solve the superposition principle, without using explicitly the asymptotics.\\
To summarize, let us briefly recall the main steps of our analysis up to this point. After linearizing the interfacial equations of hydrodynamics, we have considered the evolution equations for the perturbation modes [cfr. Eq.~\eqref{perturbationfull}]. These enabled us to determine the asymptotics at the plate [cfr. Eq.~\eqref{asymplate}] as well as the asymptotics at the bath [cfr. Eq.~\eqref{asymptbath}]. In the asymptotics at the plate we have identified the response function ${\mathcal C}$: this must be computed in order that the asymptotics at the bath confirm the boundary conditions. This selection is carried out numerically~\cite{snoeijer_2007b}. In the next section we will illustrate the main results pertaining the behaviour of the response function in terms of $q$, $\Omega$, ${\rm Ca}$ and $\theta_0$.

\section{Response function}\label{sec:response}
\subsection{A simple geometrical framework}
\begin{figure}[t!]
\includegraphics{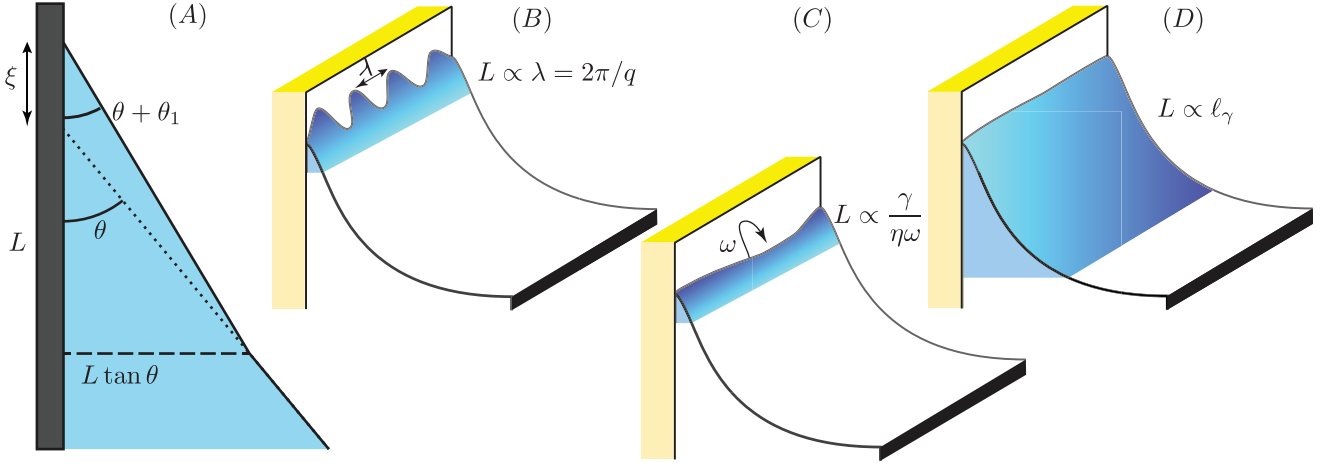}
\caption{(A) Schematic of the geometrical model relating the response function to the penetration length $L$. The length $L$ is set by the smallest of three lengths: (B) the wavelength $\lambda=2\pi/q$, (C) the dynamic length $\gamma/\eta \omega$ and (D) the capillary length $\ell_\gamma$.}
\label{Fig2}
\end{figure}
We have integrated numerically the equations derived in the previous section. To discuss the results, we wish first to propose a simple interpretation framework that will provide the scaling laws obeyed by the response function $\mathcal{C}$. Let us assume that a perturbation induced at the contact line, in time and/or in space, disturbs the interface over a penetration length $L$ along the plate. For simplicity, we consider a wedge of effective angle $\theta$. From the simple geometrical construction shown in Fig.~\ref{Fig2}A, we get for the real part of $\mathcal{C}$, at the linear order in $\hat \xi$:
\begin{equation}
R_e(\mathcal{C}) \hat \xi \equiv \cos(\theta + \theta_1) - \cos\theta \simeq \sin^2 \theta \cos \theta \frac{\hat \xi}{L}.
\end{equation}
The viscous force results from the integral over the horizontal direction of the viscous stress. Considering that the fluid moves at the same velocity as the contact line over the wedge region, we obtain the imaginary part of $\mathcal{C}$:
\begin{equation}
I_m(\mathcal{C}) \hat \xi \propto \frac{3\omega \hat \xi \eta}{\gamma \tan \theta} \ln\left(1+\frac{L \tan \theta_0}{3 \ell_s}\right).
\end{equation}
We therefore predict a relation of the form:
\begin{equation}
R_e(\mathcal{C})  \simeq \sin^2 \theta \cos \theta \frac{1}{L}\quad {\rm and} \quad I_m(\mathcal{C})  \sim \frac{3\omega \eta}{\gamma \tan \theta} \ln\left(1+\frac{L \tan \theta_0}{3 \ell_s}\right)
\end{equation}
Using the capillary length $\ell_\gamma$ to rescale $\mathcal C$, we get:
\begin{equation}
\ell_\gamma R_e(\mathcal{C})  \simeq \sin^2 \theta \cos \theta \frac{\ell_\gamma}{L}\quad {\rm and} \quad \ell_\gamma I_m(\mathcal{C})  \sim \frac{3\Omega}{\tan \theta} \ln\left(1+\frac{L \tan \theta_0}{3 \ell_s}\right).
\end{equation}
In these scaling laws, $\theta$ can be considered as the average angle at the scale $L$, which must scale according to Cox-Voinov law~\cite{COXVOINOV}:
\begin{equation}
\theta^3 \sim \theta_0^3+9 {\rm Ca} \ln{\left(1+\frac{L \tan \theta_0}{3 \ell_s}\right)}.
\end{equation}
The logarithmic factor involves the inner cut-off associated with the slip length, which explains that it involves the contact angle $\theta_0$ and not the large scale angle $\theta$. \\
The penetration length $L$ depends on three lengths that determine three asymptotic regimes, detailed in the next sections: the perturbation wavelength $\lambda=2\pi/q$ (Fig.~\ref{Fig2}B), the dynamical length $\gamma/\eta \omega=\ell_\gamma/\Omega$ set by the balance between capillary and viscous effects (Fig.~\ref{Fig2}C), and the capillary length $\ell_\gamma$, which is the outer length of the problem (Fig.~\ref{Fig2}D). 
\subsection{Dependance on $q$}
Consider an flat interface which makes an angle $\theta$ with the substrate, whose contact line is disturbed with a mode of wavenumber $q$. In static conditions, the curvature vanishes so that the interface elevation profile decays as $\sim e^{-|q| x/\cos \theta}$. The disturbance decays exponentially over a penetration length $L=|q|^{-1}\cos \theta$ along the normal $x$ to the contact line. For the real part of the response function (restoring force) we find
\begin{equation}
\gamma R_e(\mathcal{C})  \hat \xi=\gamma \sin^2\theta |q|  \hat \xi.
\end{equation}
Hence, we obtain:
\begin{equation}
\ell_\gamma R_e(\mathcal{C}) =\sin^2\theta |q|\ell_\gamma \quad {\rm and}\quad \ell_\gamma I_m(\mathcal{C})  \sim \frac{3\Omega}{\tan \theta} \ln\left(1+\frac{\cos \theta \tan \theta_0}{3 q \ell_s}\right).
\label{EqLimitedLambda}
\end{equation}
\begin{figure}[t!]
\includegraphics{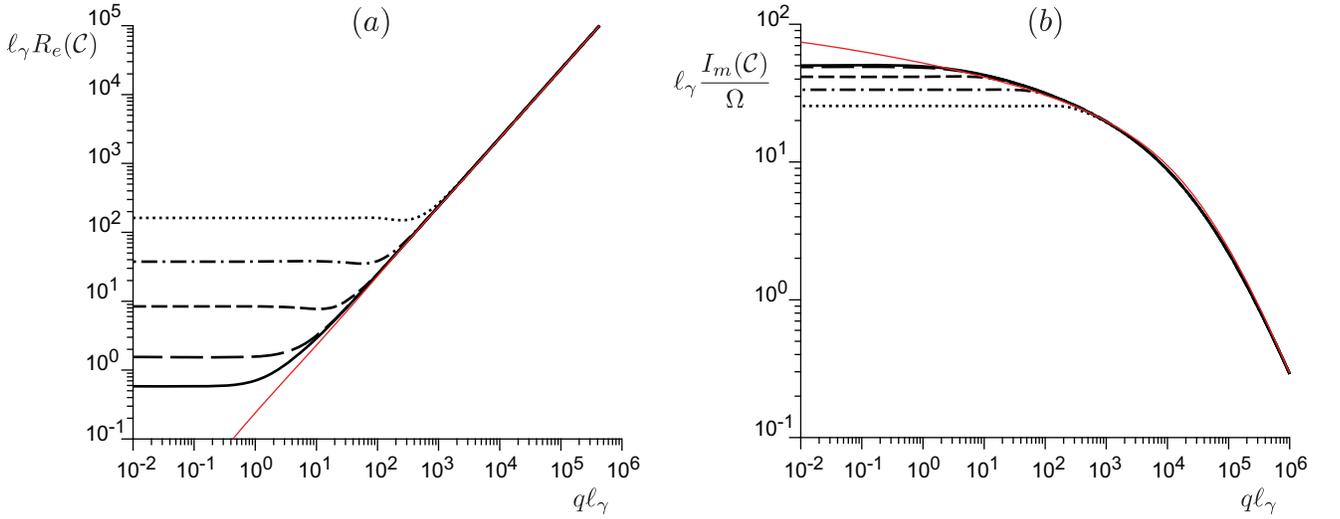}
\caption{Response function as a function of the rescaled wavenumber $q$, for different values of $\Omega$: $\Omega=0$ (solid line), $\Omega=0.22$ (long dashed line), $\Omega=1$ (dashed line), $\Omega=4.5$ (dotted dashed line), $\Omega=20$ (dotted line). The other parameters are fixed at $\tan \theta_0=0.55$, $\ell_s=2.5\,10^{-6}\ell_\gamma$ and ${\rm Ca}=10^{-5}$. The thin red lines are the predictions for the large $q$ asymptotics given by Eq.~\eqref{EqLimitedLambda}, in which a corrective factor $0.87$ was applied in front of the imaginary part.}
\label{Fig4}
\end{figure}
\begin{figure}[t!]
\includegraphics{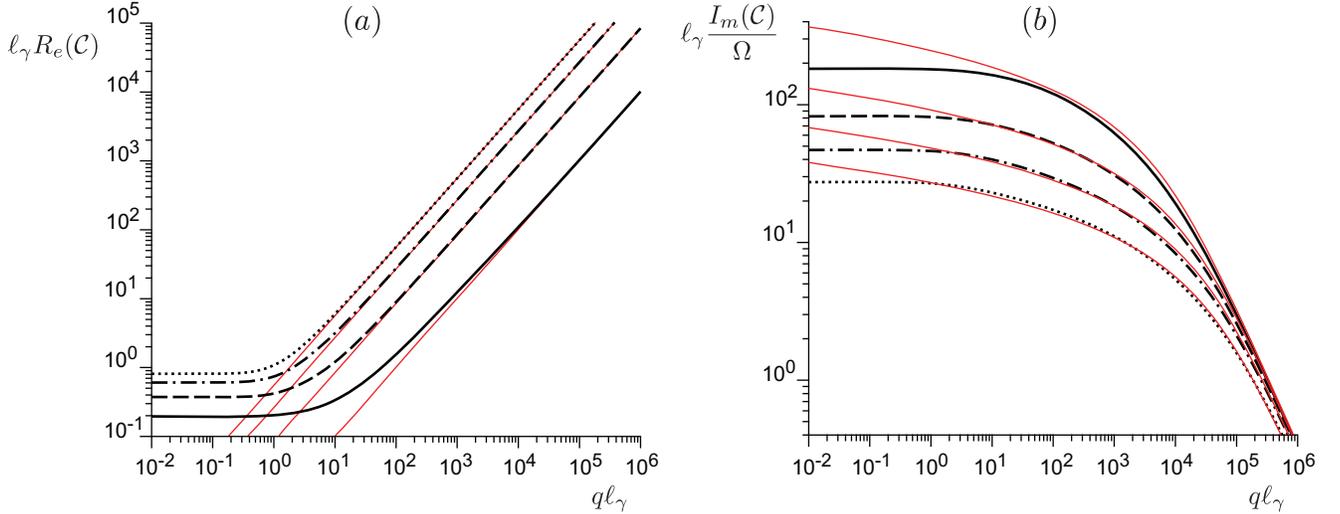}
\caption{Response function as a function of the rescaled wavenumber $q$, in the limit of vanishing $\Omega$, for different angles $\tan \theta_0$: $\tan \theta_0=0.1$ (solid line), $\tan \theta_0=0.3$ (dashed line), $\tan \theta_0=0.6$ (dotted dashed line) and $\tan \theta_0=1.1$ (dotted line). The other parameters are fixed at $\ell_s=2.5\,10^{-6}\ell_\gamma$ and ${\rm Ca}=10^{-5}$. The thin red lines are the predictions for the large $q$ asymptotics given by Eq.~\eqref{EqLimitedLambda}, in which a corrective factor $0.87$ was applied in front of the imaginary part.}
\label{Fig5}
\end{figure}
\begin{figure}[t!]
\includegraphics{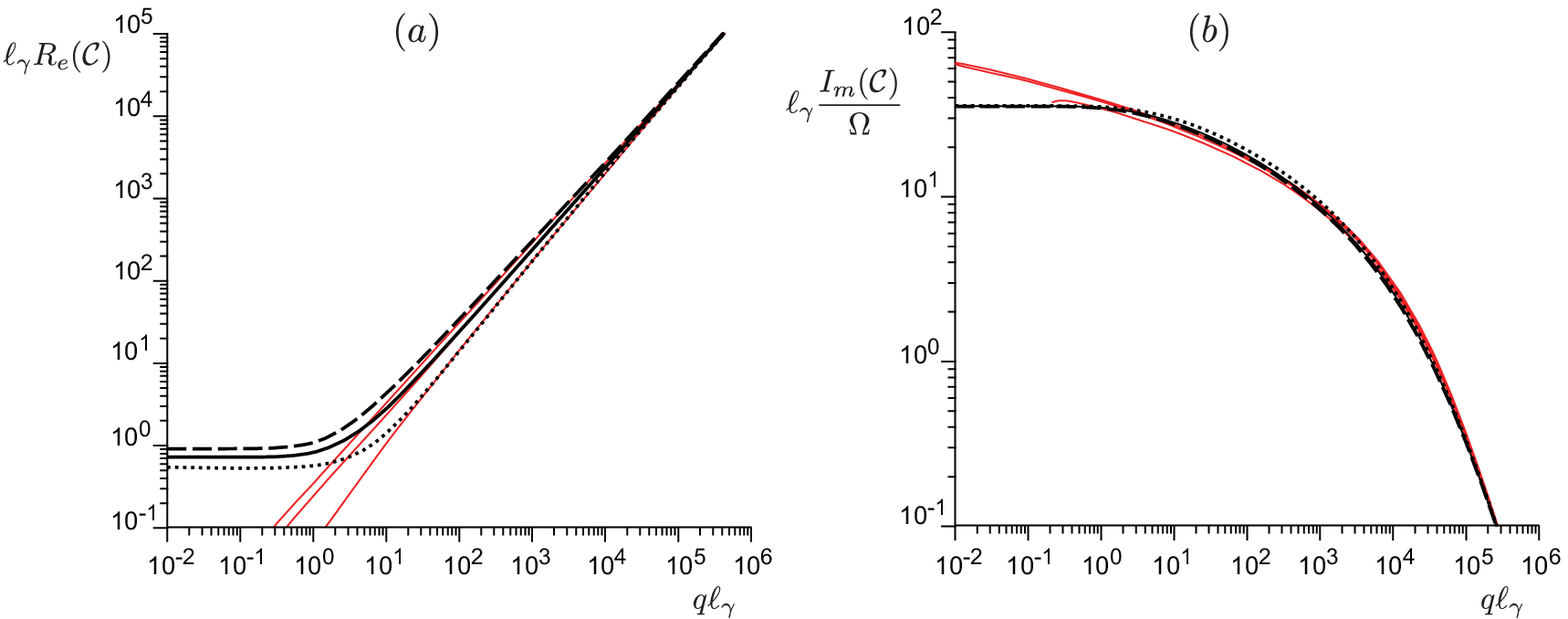}
\caption{Response function as a function of the rescaled wavenumber $q$, in the limit of vanishing $\Omega$, for different values of $\rm Ca$: $\rm Ca=-10^{-3}$ (dotted line), $\rm Ca=0$ (solid line), $\rm Ca=10^{-3}$ (dashed line). The other parameters are fixed at $\tan \theta_0=0.55$, $\ell_s=2.5\,10^{-6}\ell_\gamma$.The thin red lines are the predictions for the large $q$ asymptotics given by Eq.~\eqref{EqLimitedLambda}, in which a corrective factor $0.87$ was applied in front of the imaginary part.}
\label{Fig7}
\end{figure}
Figure~\ref{Fig4} shows the dependence of $\mathcal{C}$ with respect to $q$ for different values of $\Omega$. One observes that the large $q$ regime is independent of $\Omega$ (provided $\Omega$ is small enough) and nicely coincides with the prediction of a quasi-static disturbance at vanishing curvature. The prediction for $R_e(\mathcal{C})$ is quantitative but that for $I_m(\mathcal{C})$ is overestimated by $\simeq 13\%$. A multiplicative factor $0.87$ was accordingly applied when plotting the predictions. As expected, $\mathcal{C}$  strongly depends on the (true) contact angle $\theta_0$ in this large $q$ regime (Fig.~\ref{Fig5}). This dependence is quantitatively predicted by Eq.~\eqref{EqLimitedLambda}. Finally, Fig.~\ref{Fig7} shows that there is a small dependence on the capillary number $\rm Ca$ in this regime, that we interpret as resulting from the change of the interface slope with the scale $q^{-1}$.  As predicted by Eq.~\eqref{EqLimitedLambda}, the dependence is weak for $I_m(\mathcal{C})$ as the dependence on $L$ is logarithmic, and larger for $R_e(\mathcal{C})$, which linearly depends on $L^{-1}$. Given the crude assumptions made in the geometrical model, the excellent agreement validates this interpretation.

\subsection{Dependence on $\Omega$}
\begin{figure}[t!]
\includegraphics{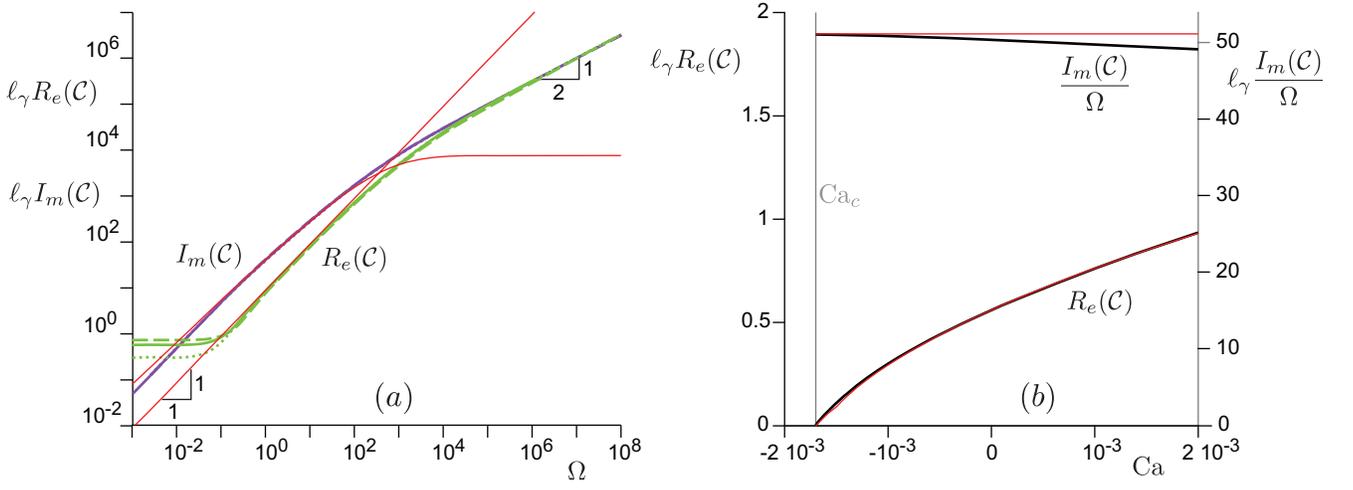}
\caption{Left Panel: response function as a function of the rescaled frequency $\Omega$, in the limit of vanishing $q$, for different values of $\rm Ca$: $\rm Ca=-10^{-3}$ (dotted line), $\rm Ca=0$ (solid line), $\rm Ca=10^{-3}$ (dashed line). The other parameters are fixed at $\tan \theta_0=0.55$, $\ell_s=2.5\,10^{-6}\ell_\gamma$. The three curves almost collapse for the imaginary part of $\mathcal C$ (blue curves) but are spitted for its real part (green curves), at small $\Omega$. The thin red lines are the predictions of the intermediate asymptotics limited by the dynamical length (small $q$ limit and intermediate range of $\Omega$), as given by Eq.~\eqref{EqLimitedOmega}. Right Panel: response function in the limit of vanishing $\Omega$ and vanishing $q$, as a function of $\rm Ca$. The thin red lines are the predictions of the quasi-steady asymptotics (small $q$ limit and small $\Omega$), as given by Eq.~\eqref{EqLimitedEllGamma}.}
\label{Fig6}
\end{figure}

Figure~\ref{Fig4} shows that there is a cross-over at small $q$ between the large $q$ regime discussed above and a regime which depends on the frequency $\Omega$, but not on $q$. Figure~\ref{Fig6} shows the dependence on $\Omega$ in this small $q$ limit, which presents three asymptotics. In the limit of vanishing $\Omega$, one observes a plateau of $R_e(\mathcal{C})$ while $I_m(\mathcal{C})$ is linear in $\Omega$. This quasi-steady asymptotics is discussed in the next section. In the large $\Omega$ limit, one observes a power law asymptotics $R_e(\mathcal{C})=I_m(\mathcal{C})\propto \Omega^{1/2}$. Finally, in the intermediate asymptotics in $\Omega$, $R_e(\mathcal{C})$ appears to be linear in $\Omega$ while $I_m(\mathcal{C})$ is sublinear. This corresponds to the dynamical regime that we now describe, based on the geometrical argument schematized in Fig.~\ref{Fig2}C. In the limit of intermediate $\Omega$, the modes penetrate on the interface over a length smaller than the capillary length $\ell_\gamma$ but larger than $\ell_s$. In order to determine the penetration length $L$ in this case, we can therefore replace the shape of the interface by a wedge at angle $\theta$ writing $h_0'=\tan \theta$, which leads to the equation:
\begin{equation}
 (x^3 h'''_1)' =  - j \frac{3 \Omega }{\ell_\gamma \sin^3 \theta}h_1.
\end{equation}
The above equation presents two independent solutions that converge at infinity (far from the contact line), based on the $\mbox{MeijerG}$ special function~\cite{GR}. Here, we just identify the length $L$, in its dimensional form, as:
\begin{equation}
L\sim \frac{\sin^3\theta  \ell_\gamma}{3 \Omega }.
\end{equation}
A refined asymptotic treatment gives the multiplicative constants involved in front of $L$:
\begin{equation}
R_e(\mathcal{C})  \simeq \frac{3\pi\omega \eta}{2 \gamma \tan \theta} \quad {\rm and} \quad I_m(\mathcal{C})  \sim \frac{3\omega \eta }{\gamma \tan \theta} \ln\left(1+\frac{\tan\theta_0 \sin^3\theta_0 \ell_\gamma}{3 \exp(4 \gamma_{\mbox{\tiny Euler}} -\frac{1}{2}) \Omega \ell_s}\right)
\label{EqLimitedOmega}
\end{equation}
where the slip length $\ell_s$ has here the dimension of a length and $\gamma_{\mbox{\tiny Euler}} \simeq 0.577\cdot$ is the Euler-Mascheroni constant. Rescaling $\mathcal C$ by the capillary length, we get:
\begin{equation}
\ell_\gamma R_e(\mathcal{C})  \simeq \frac{3\pi\Omega}{2\tan \theta} \quad {\rm and} \quad \ell_\gamma I_m(\mathcal{C})  \sim \frac{3\Omega }{ \tan \theta} \ln\left(1+\frac{\tan\theta_0 \sin^3\theta_0 \ell_\gamma}{3 \exp(4 \gamma_{\mbox{\tiny Euler}} -\frac{1}{2}) \Omega \ell_s}\right).
\label{EqLimitedOmega}
\end{equation}
The left panel of Fig.~\ref{Fig6} shows that the agreement of this model equation with the exact response function is, again, very good in the intermediate range of $\Omega$. The failure appears at large $\Omega$, when the penetration length $L$ reaches the slip length $\ell_s$. The large $\Omega$ asymptotics is therefore a bit artificial, as the penetration length becomes comparable to the molecular size. This asymptotics is therefore sensitive to the details of the modeling. In order to keep the clarity of the paper, we will let it apart here. Note that the observed asymptotics $R_e(\mathcal{C})=I_m(\mathcal{C})\propto \Omega^{1/2}$ is solution of the Kramers-Kronig relation. At small $\Omega$, the cross-over appears when the dynamical length $L$ reaches the outer length $\ell_\gamma$, which sets the size of the meniscus. We discuss this asymptotic in the next section.


\subsection{Dependence on ${\rm Ca}$}\label{sec:dependenceonCa}
In the double limit of vanishing $\Omega$ and vanishing $q$, the time evolution is slow and the wavelength much larger than the capillary length. In first approximation, moving the contact line with respect to the plate is the same as moving the plate with respect to the contact line. The main difference lies in the dissipation around the bath. We therefore assume that the dependence of the equilibrium altitude $\delta$ with respect to $\rm Ca$ and $\theta_0$ holds during transients, provided one replaces $\delta$ by $\delta+\xi$ and $\rm Ca$ by ${\rm Ca}+d\xi/dt$ in these laws:
\begin{eqnarray}
\delta &=&\sqrt{2(1-\sin \theta_M)}\,\ell_\gamma\\
\theta_M^3 &=& \theta_0^3+9 {\rm Ca}\; \ln{\left(\frac{ \alpha \ell_\gamma}{3\ell_s}\right)}.
\end{eqnarray}
Differentiating these expressions with respect to $\theta_0$, $\delta$ and $\rm Ca$, one obtains:
\begin{equation}
\frac{\theta_0^2}{\sin \theta_0} \ell_\gamma {\mathcal C}=\frac{\theta_M^2\sqrt{2(1-\sin \theta_M)}}{\cos \theta_M}+3j \Omega  \ln{\left(\frac{ \alpha \ell_\gamma}{3\ell_s}\right)}
\label{EqLimitedEllGamma}
\end{equation}
where the slip length $\ell_s$ is not rescaled by $\ell_\gamma$. Note that in this expression, we ignored the small dependencies of $\alpha$ with $\theta_0$. The right panel of Fig.~\ref{Fig6} shows that the agreement is perfect for the real part of ${\mathcal C}$, as the quasi-steady limit is an exact, controlled approximation of the problem for the restoring force. The agreement of the imaginary part of ${\mathcal C}$, which is linear in $\Omega$, is good but not perfect. The approximation used assumes that it is equivalent to impose a displacement of the contact line with respect to the solid plate (the bath remaining fixed) and to impose a displacement of the solid plate with respect to the contact line and the bath. If the small scales of the problem are indeed equivalent and are entirely determined by the relative displacement of the contact line with respect to the plate, the large scales differ if the contact line or the plate move with respect to the bath. The small difference between the prediction and the exact  ${\mathcal C}$ results from the small dissipation at the scale of the meniscus.

\section{Concluding remarks}\label{sec:concluding}
The motion of a contact line on a nanoscale random heterogeneous landscape is one of the most important problems remaining partly open in the field of dynamical wetting~\cite{bonn_wetting_2009,snoeijer_moving_2013,kwon2013increasing,sirringhaus2000high,van2008inkjet,ahn2009omnidirectional,onses2013hierarchical,bird2013reducing,liu2014pancake,fan2000rapid,park2007high,stuart2010emerging,duprat2012wetting,galliker2012direct,paxson2013self}. Most applications may require only the knowledge of elementary information, e.g. the time evolution of the average contact line position, rather than the complete details of the full problem. In this framework, working in the widely used set-up of a plate withdrawal from a bath at a constant velocity~\cite{E04b,E05,eggers_contact_2005,ChanPOF12}, we have derived a simplified force balance equation governing the time-evolution of the contact line. The key steps of the analysis can be briefly summarized as follows: first, we have treated the problem at the hydrodynamic scales, based on the lubrication approximation~\cite{ODB97}. Second, the space-time evolution of defects perturbations have been deduced by linearizing the lubrication equations around a base state~\cite{snoeijer_2007b}, followed by double Fourier transform in space and time. Finally, the effects of perturbations can be averaged along the contact line position to design a simplified deterministic model for the evolution of the contact line [cfr. Eq.~\eqref{eq:deterministic_eq}]. Our work hinges on the characterization of the {\it response function} of the contact line, relating its displacement to the forces set by heterogeneities [cfr. Eqs.~\eqref{psieq}-\eqref{eq:Cdecomposition}]. These are the restoring force associated with the elastic deformation of the liquid interface, which appears as a memory term in the model, and the force set by the spatial variation of the contact angle.\\
Once the simplified deterministic model is obtained, several applications can be envisaged. Primarily, we have illustrated how to obtain a stochastic model accounting for thermal fluctuations [cfr. Eq.~\eqref{eq:Langevin}], which allows to study numerically the activated motion of the contact line through the frozen energy landscape provided by defects. The equations involve a memory term, hence numerical simulations of this model require the storage of the contact line profile as a function of time, limiting the possibility for current computers to $3$ or $4$ decades in space. A first application would be to perform a side-by-side comparisons between the outcome of experiments and the prediction of the stochastic model. This requires the experimental determination of the energy landscape of a well controlled heterogeneous substrate and experimental visualization of the contact line at the nanoscales. As the model is derived in the linear response regime, this opens the question, beside technical difficulties, of possible non-linear effects. We notice that the effects of thermal fluctuations on the interfacial equations of hydrodynamics should be consistently accompanied by the characterization of some molecular features close to the wall (e.g. disjoining pressure~\cite{eggers_contact_2005}). In this perspective, as already remarked in the text, we notice that expression \eqref{eq:noise_form} is expected to be valid within a proportionality coefficient that depends on the ``inner layer'' of the problem, hence on the detailed molecular interactions and their coupling with thermal fluctuations close to the wall, as explained in~\cite{belardinelli2016thermal}. Such prefactor is unknown at this level of description, but a useful parametrization has been discussed in~\cite{belardinelli2016thermal}.\\
The potentiality of the approach presented in this paper takes somehow a broader perspective than just the force balance equation presented in Eq.~\eqref{eq:Langevin}. Regarding thermal fluctuations, for example, such equation allows for an easy mathematical generalization in presence of an external driving mechanism periodically oscillating in time. This can be used to analyze the phenomenon of stochastic resonance~\cite{Stochastic1,Stochastic2}, by studying the interplay between the characteristic hopping time scale set by thermal activation and the external time scale set by the periodic driving. Possibly, this could help to design appropriate experiments to give further credit to the thermally activated scenario~\cite{Prevost99,perrin_2016}.\\ 
Finally, we wish to remark that the framework derived here opens the promising perspective of treating the mechanical behaviour of a contact line, its "rheology", using low-dimensional models~\cite{Martens,Mansard,Picard,HL98,Sollich1,Sollich2,Sollich3}. It directly connects to current progresses in soft-glassy materials~\cite{BonnReviewYIELDSTRESS}, which shares strong similarities with the contact line problem: the multi-scale character, the presence of a dynamic critical point, a self-built energy landscape. Controlled reductions to low-dimensional models may help to overcome the intrinsic limit of the formulation derived here (linear response; a well-known large-scale asymptotics). A potential application is to solve the inverse problem and to determine mechanical/chemical properties of an interface using a contact line. Such a contact line nano-rheometer would be particularly interesting in the case of soft solid, with direct applications on bio-medical tissues. \\

H.P. and D.B. have equally contributed to this work. B.A. is supported by Institut Universitaire de France. This work was funded by the ANR grant Smart.

\bibliography{PrFluidsLubrifCL}

\begin{thebibliography}{60}
\expandafter\ifx\csname natexlab\endcsname\relax\def\natexlab#1{#1}\fi
\expandafter\ifx\csname bibnamefont\endcsname\relax
  \def\bibnamefont#1{#1}\fi
\expandafter\ifx\csname bibfnamefont\endcsname\relax
  \def\bibfnamefont#1{#1}\fi
\expandafter\ifx\csname citenamefont\endcsname\relax
  \def\citenamefont#1{#1}\fi
\expandafter\ifx\csname url\endcsname\relax
  \def\url#1{\texttt{#1}}\fi
\expandafter\ifx\csname urlprefix\endcsname\relax\def\urlprefix{URL }\fi
\providecommand{\bibinfo}[2]{#2}
\providecommand{\eprint}[2][]{\url{#2}}

\bibitem[{\citenamefont{Bonn et~al.}(2009)\citenamefont{Bonn, Eggers, Indekeu,
  Meunier, and Rolley}}]{bonn_wetting_2009}
\bibinfo{author}{\bibfnamefont{D.}~\bibnamefont{Bonn}},
  \bibinfo{author}{\bibfnamefont{J.}~\bibnamefont{Eggers}},
  \bibinfo{author}{\bibfnamefont{J.}~\bibnamefont{Indekeu}},
  \bibinfo{author}{\bibfnamefont{J.}~\bibnamefont{Meunier}}, \bibnamefont{and}
  \bibinfo{author}{\bibfnamefont{E.}~\bibnamefont{Rolley}},
  \bibinfo{journal}{Rev. Mod. Phys.} \textbf{\bibinfo{volume}{81}},
  \bibinfo{pages}{739} (\bibinfo{year}{2009}).

\bibitem[{\citenamefont{Snoeijer and Andreotti}(2013)}]{snoeijer_moving_2013}
\bibinfo{author}{\bibfnamefont{J.~H.} \bibnamefont{Snoeijer}} \bibnamefont{and}
  \bibinfo{author}{\bibfnamefont{B.}~\bibnamefont{Andreotti}},
  \bibinfo{journal}{Annu. Rev. Fluid Mech.} \textbf{\bibinfo{volume}{45}},
  \bibinfo{pages}{269} (\bibinfo{year}{2013}).

\bibitem[{\citenamefont{Wang et~al.}(2014)\citenamefont{Wang, Do-Quang, Cannon,
  Yue, Suzuki, Amberg, and Shiomi}}]{Amberg14}
\bibinfo{author}{\bibfnamefont{J.}~\bibnamefont{Wang}},
  \bibinfo{author}{\bibfnamefont{M.}~\bibnamefont{Do-Quang}},
  \bibinfo{author}{\bibfnamefont{J.~J.} \bibnamefont{Cannon}},
  \bibinfo{author}{\bibfnamefont{f.}~\bibnamefont{Yue}},
  \bibinfo{author}{\bibfnamefont{Y.}~\bibnamefont{Suzuki}},
  \bibinfo{author}{\bibfnamefont{G.}~\bibnamefont{Amberg}}, \bibnamefont{and}
  \bibinfo{author}{\bibfnamefont{J.}~\bibnamefont{Shiomi}},
  \bibinfo{journal}{Scientific Reports} \textbf{\bibinfo{volume}{5}},
  \bibinfo{pages}{8474} (\bibinfo{year}{2014}).

\bibitem[{\citenamefont{Kwon et~al.}(2013)\citenamefont{Kwon, Bird, and
  Varanasi}}]{kwon2013increasing}
\bibinfo{author}{\bibfnamefont{H.-m.} \bibnamefont{Kwon}},
  \bibinfo{author}{\bibfnamefont{J.~C.} \bibnamefont{Bird}}, \bibnamefont{and}
  \bibinfo{author}{\bibfnamefont{K.~K.} \bibnamefont{Varanasi}},
  \bibinfo{journal}{Applied Physics Letters} \textbf{\bibinfo{volume}{103}},
  \bibinfo{pages}{201601} (\bibinfo{year}{2013}).

\bibitem[{\citenamefont{Sirringhaus et~al.}(2000)\citenamefont{Sirringhaus,
  Kawase, Friend, Shimoda, Inbasekaran, Wu, and Woo}}]{sirringhaus2000high}
\bibinfo{author}{\bibfnamefont{H.}~\bibnamefont{Sirringhaus}},
  \bibinfo{author}{\bibfnamefont{T.}~\bibnamefont{Kawase}},
  \bibinfo{author}{\bibfnamefont{R.}~\bibnamefont{Friend}},
  \bibinfo{author}{\bibfnamefont{T.}~\bibnamefont{Shimoda}},
  \bibinfo{author}{\bibfnamefont{M.}~\bibnamefont{Inbasekaran}},
  \bibinfo{author}{\bibfnamefont{W.}~\bibnamefont{Wu}}, \bibnamefont{and}
  \bibinfo{author}{\bibfnamefont{E.}~\bibnamefont{Woo}},
  \bibinfo{journal}{Science} \textbf{\bibinfo{volume}{290}},
  \bibinfo{pages}{2123} (\bibinfo{year}{2000}).

\bibitem[{\citenamefont{Van~Osch et~al.}(2008)\citenamefont{Van~Osch, Perelaer,
  de~Laat, and Schubert}}]{van2008inkjet}
\bibinfo{author}{\bibfnamefont{T.~H.} \bibnamefont{Van~Osch}},
  \bibinfo{author}{\bibfnamefont{J.}~\bibnamefont{Perelaer}},
  \bibinfo{author}{\bibfnamefont{A.~W.} \bibnamefont{de~Laat}},
  \bibnamefont{and} \bibinfo{author}{\bibfnamefont{U.~S.}
  \bibnamefont{Schubert}}, \bibinfo{journal}{Advanced Materials}
  \textbf{\bibinfo{volume}{20}}, \bibinfo{pages}{343} (\bibinfo{year}{2008}).

\bibitem[{\citenamefont{Ahn et~al.}(2009)\citenamefont{Ahn, Duoss, Motala, Guo,
  Park, Xiong, Yoon, Nuzzo, Rogers, and Lewis}}]{ahn2009omnidirectional}
\bibinfo{author}{\bibfnamefont{B.~Y.} \bibnamefont{Ahn}},
  \bibinfo{author}{\bibfnamefont{E.~B.} \bibnamefont{Duoss}},
  \bibinfo{author}{\bibfnamefont{M.~J.} \bibnamefont{Motala}},
  \bibinfo{author}{\bibfnamefont{X.}~\bibnamefont{Guo}},
  \bibinfo{author}{\bibfnamefont{S.-I.} \bibnamefont{Park}},
  \bibinfo{author}{\bibfnamefont{Y.}~\bibnamefont{Xiong}},
  \bibinfo{author}{\bibfnamefont{J.}~\bibnamefont{Yoon}},
  \bibinfo{author}{\bibfnamefont{R.~G.} \bibnamefont{Nuzzo}},
  \bibinfo{author}{\bibfnamefont{J.~A.} \bibnamefont{Rogers}},
  \bibnamefont{and} \bibinfo{author}{\bibfnamefont{J.~A.} \bibnamefont{Lewis}},
  \bibinfo{journal}{Science} \textbf{\bibinfo{volume}{323}},
  \bibinfo{pages}{1590} (\bibinfo{year}{2009}).

\bibitem[{\citenamefont{Onses et~al.}(2013)\citenamefont{Onses, Song,
  Williamson, Sutanto, Ferreira, Alleyne, Nealey, Ahn, and
  Rogers}}]{onses2013hierarchical}
\bibinfo{author}{\bibfnamefont{M.~S.} \bibnamefont{Onses}},
  \bibinfo{author}{\bibfnamefont{C.}~\bibnamefont{Song}},
  \bibinfo{author}{\bibfnamefont{L.}~\bibnamefont{Williamson}},
  \bibinfo{author}{\bibfnamefont{E.}~\bibnamefont{Sutanto}},
  \bibinfo{author}{\bibfnamefont{P.~M.} \bibnamefont{Ferreira}},
  \bibinfo{author}{\bibfnamefont{A.~G.} \bibnamefont{Alleyne}},
  \bibinfo{author}{\bibfnamefont{P.~F.} \bibnamefont{Nealey}},
  \bibinfo{author}{\bibfnamefont{H.}~\bibnamefont{Ahn}}, \bibnamefont{and}
  \bibinfo{author}{\bibfnamefont{J.~A.} \bibnamefont{Rogers}},
  \bibinfo{journal}{Nature Nanotechnology} \textbf{\bibinfo{volume}{8}},
  \bibinfo{pages}{667} (\bibinfo{year}{2013}).

\bibitem[{\citenamefont{Bird et~al.}(2013)\citenamefont{Bird, Dhiman, Kwon, and
  Varanasi}}]{bird2013reducing}
\bibinfo{author}{\bibfnamefont{J.~C.} \bibnamefont{Bird}},
  \bibinfo{author}{\bibfnamefont{R.}~\bibnamefont{Dhiman}},
  \bibinfo{author}{\bibfnamefont{H.-M.} \bibnamefont{Kwon}}, \bibnamefont{and}
  \bibinfo{author}{\bibfnamefont{K.~K.} \bibnamefont{Varanasi}},
  \bibinfo{journal}{Nature} \textbf{\bibinfo{volume}{503}},
  \bibinfo{pages}{385} (\bibinfo{year}{2013}).

\bibitem[{\citenamefont{Liu et~al.}(2014)\citenamefont{Liu, Moevius, Xu, Qian,
  Yeomans, and Wang}}]{liu2014pancake}
\bibinfo{author}{\bibfnamefont{Y.}~\bibnamefont{Liu}},
  \bibinfo{author}{\bibfnamefont{L.}~\bibnamefont{Moevius}},
  \bibinfo{author}{\bibfnamefont{X.}~\bibnamefont{Xu}},
  \bibinfo{author}{\bibfnamefont{T.}~\bibnamefont{Qian}},
  \bibinfo{author}{\bibfnamefont{J.~M.} \bibnamefont{Yeomans}},
  \bibnamefont{and} \bibinfo{author}{\bibfnamefont{Z.}~\bibnamefont{Wang}},
  \bibinfo{journal}{Nature Physics} \textbf{\bibinfo{volume}{10}},
  \bibinfo{pages}{515} (\bibinfo{year}{2014}).

\bibitem[{\citenamefont{Fan et~al.}(2000)\citenamefont{Fan, Lu, Stump, Reed,
  Baer, Schunk, Perez-Luna, L{\'o}pez, and Brinker}}]{fan2000rapid}
\bibinfo{author}{\bibfnamefont{H.}~\bibnamefont{Fan}},
  \bibinfo{author}{\bibfnamefont{Y.}~\bibnamefont{Lu}},
  \bibinfo{author}{\bibfnamefont{A.}~\bibnamefont{Stump}},
  \bibinfo{author}{\bibfnamefont{S.~T.} \bibnamefont{Reed}},
  \bibinfo{author}{\bibfnamefont{T.}~\bibnamefont{Baer}},
  \bibinfo{author}{\bibfnamefont{R.}~\bibnamefont{Schunk}},
  \bibinfo{author}{\bibfnamefont{V.}~\bibnamefont{Perez-Luna}},
  \bibinfo{author}{\bibfnamefont{G.~P.} \bibnamefont{L{\'o}pez}},
  \bibnamefont{and} \bibinfo{author}{\bibfnamefont{C.~J.}
  \bibnamefont{Brinker}}, \bibinfo{journal}{Nature}
  \textbf{\bibinfo{volume}{405}}, \bibinfo{pages}{56} (\bibinfo{year}{2000}).

\bibitem[{\citenamefont{Park et~al.}(2007)\citenamefont{Park, Hardy, Kang,
  Barton, Adair, Kishore~Mukhopadhyay, Lee, Strano, Alleyne, Georgiadis
  et~al.}}]{park2007high}
\bibinfo{author}{\bibfnamefont{J.-U.} \bibnamefont{Park}},
  \bibinfo{author}{\bibfnamefont{M.}~\bibnamefont{Hardy}},
  \bibinfo{author}{\bibfnamefont{S.~J.} \bibnamefont{Kang}},
  \bibinfo{author}{\bibfnamefont{K.}~\bibnamefont{Barton}},
  \bibinfo{author}{\bibfnamefont{K.}~\bibnamefont{Adair}},
  \bibinfo{author}{\bibfnamefont{D.}~\bibnamefont{Kishore~Mukhopadhyay}},
  \bibinfo{author}{\bibfnamefont{C.~Y.} \bibnamefont{Lee}},
  \bibinfo{author}{\bibfnamefont{M.~S.} \bibnamefont{Strano}},
  \bibinfo{author}{\bibfnamefont{A.~G.} \bibnamefont{Alleyne}},
  \bibinfo{author}{\bibfnamefont{J.~G.} \bibnamefont{Georgiadis}},
  \bibnamefont{et~al.}, \bibinfo{journal}{Nature Materials}
  \textbf{\bibinfo{volume}{6}}, \bibinfo{pages}{782} (\bibinfo{year}{2007}).

\bibitem[{\citenamefont{Stuart et~al.}(2010)\citenamefont{Stuart, Huck, Genzer,
  M{\"u}ller, Ober, Stamm, Sukhorukov, Szleifer, Tsukruk, Urban
  et~al.}}]{stuart2010emerging}
\bibinfo{author}{\bibfnamefont{M.~A.~C.} \bibnamefont{Stuart}},
  \bibinfo{author}{\bibfnamefont{W.~T.} \bibnamefont{Huck}},
  \bibinfo{author}{\bibfnamefont{J.}~\bibnamefont{Genzer}},
  \bibinfo{author}{\bibfnamefont{M.}~\bibnamefont{M{\"u}ller}},
  \bibinfo{author}{\bibfnamefont{C.}~\bibnamefont{Ober}},
  \bibinfo{author}{\bibfnamefont{M.}~\bibnamefont{Stamm}},
  \bibinfo{author}{\bibfnamefont{G.~B.} \bibnamefont{Sukhorukov}},
  \bibinfo{author}{\bibfnamefont{I.}~\bibnamefont{Szleifer}},
  \bibinfo{author}{\bibfnamefont{V.~V.} \bibnamefont{Tsukruk}},
  \bibinfo{author}{\bibfnamefont{M.}~\bibnamefont{Urban}},
  \bibnamefont{et~al.}, \bibinfo{journal}{Nature materials}
  \textbf{\bibinfo{volume}{9}}, \bibinfo{pages}{101} (\bibinfo{year}{2010}).

\bibitem[{\citenamefont{Duprat et~al.}(2012)\citenamefont{Duprat, Protiere,
  Beebe, and Stone}}]{duprat2012wetting}
\bibinfo{author}{\bibfnamefont{C.}~\bibnamefont{Duprat}},
  \bibinfo{author}{\bibfnamefont{S.}~\bibnamefont{Protiere}},
  \bibinfo{author}{\bibfnamefont{A.}~\bibnamefont{Beebe}}, \bibnamefont{and}
  \bibinfo{author}{\bibfnamefont{H.}~\bibnamefont{Stone}},
  \bibinfo{journal}{Nature} \textbf{\bibinfo{volume}{482}},
  \bibinfo{pages}{510} (\bibinfo{year}{2012}).

\bibitem[{\citenamefont{Galliker et~al.}(2012)\citenamefont{Galliker,
  Schneider, Eghlidi, Kress, Sandoghdar, and Poulikakos}}]{galliker2012direct}
\bibinfo{author}{\bibfnamefont{P.}~\bibnamefont{Galliker}},
  \bibinfo{author}{\bibfnamefont{J.}~\bibnamefont{Schneider}},
  \bibinfo{author}{\bibfnamefont{H.}~\bibnamefont{Eghlidi}},
  \bibinfo{author}{\bibfnamefont{S.}~\bibnamefont{Kress}},
  \bibinfo{author}{\bibfnamefont{V.}~\bibnamefont{Sandoghdar}},
  \bibnamefont{and}
  \bibinfo{author}{\bibfnamefont{D.}~\bibnamefont{Poulikakos}},
  \bibinfo{journal}{Nature Communications} \textbf{\bibinfo{volume}{3}},
  \bibinfo{pages}{890} (\bibinfo{year}{2012}).

\bibitem[{\citenamefont{Paxson and Varanasi}(2013)}]{paxson2013self}
\bibinfo{author}{\bibfnamefont{A.~T.} \bibnamefont{Paxson}} \bibnamefont{and}
  \bibinfo{author}{\bibfnamefont{K.~K.} \bibnamefont{Varanasi}},
  \bibinfo{journal}{Nature Communications} \textbf{\bibinfo{volume}{4}},
  \bibinfo{pages}{1492} (\bibinfo{year}{2013}).

\bibitem[{\citenamefont{Eggers}(2004)}]{E04b}
\bibinfo{author}{\bibfnamefont{J.}~\bibnamefont{Eggers}},
  \bibinfo{journal}{Phys. Rev. Lett.} \textbf{\bibinfo{volume}{93}},
  \bibinfo{pages}{094502} (\bibinfo{year}{2004}).

\bibitem[{\citenamefont{Eggers}(2005{\natexlab{a}})}]{E05}
\bibinfo{author}{\bibfnamefont{J.}~\bibnamefont{Eggers}},
  \bibinfo{journal}{Phys. Fluids} \textbf{\bibinfo{volume}{17}},
  \bibinfo{pages}{082106} (\bibinfo{year}{2005}{\natexlab{a}}).

\bibitem[{\citenamefont{Eggers}(2005{\natexlab{b}})}]{eggers_contact_2005}
\bibinfo{author}{\bibfnamefont{J.}~\bibnamefont{Eggers}},
  \bibinfo{journal}{Phys. Rev. E} \textbf{\bibinfo{volume}{72}},
  \bibinfo{pages}{061605} (\bibinfo{year}{2005}{\natexlab{b}}).

\bibitem[{\citenamefont{Chan et~al.}(2012)\citenamefont{Chan, Snoeijer, and
  Eggers}}]{ChanPOF12}
\bibinfo{author}{\bibfnamefont{T.}~\bibnamefont{Chan}},
  \bibinfo{author}{\bibfnamefont{J.}~\bibnamefont{Snoeijer}}, \bibnamefont{and}
  \bibinfo{author}{\bibfnamefont{J.}~\bibnamefont{Eggers}},
  \bibinfo{journal}{Phys. Fluids} \textbf{\bibinfo{volume}{24}},
  \bibinfo{pages}{072104} (\bibinfo{year}{2012}).

\bibitem[{\citenamefont{de~Gennes et~al.}(2002)\citenamefont{de~Gennes,
  Brochart-Wyart, and D.}}]{deGe02}
\bibinfo{author}{\bibfnamefont{P.-G.} \bibnamefont{de~Gennes}},
  \bibinfo{author}{\bibfnamefont{F.}~\bibnamefont{Brochart-Wyart}},
  \bibnamefont{and} \bibinfo{author}{\bibfnamefont{Q.}~\bibnamefont{D.}},
  \emph{\bibinfo{title}{Gouttes, bulles, perles et ondes.}}
  (\bibinfo{publisher}{Belin}, \bibinfo{year}{2002}).

\bibitem[{\citenamefont{Snoeijer and
  Andreotti}(2008)}]{snoeijer_microscopic_2008}
\bibinfo{author}{\bibfnamefont{J.~H.} \bibnamefont{Snoeijer}} \bibnamefont{and}
  \bibinfo{author}{\bibfnamefont{B.}~\bibnamefont{Andreotti}},
  \bibinfo{journal}{Phys. Fluids} \textbf{\bibinfo{volume}{20}},
  \bibinfo{pages}{057101} (\bibinfo{year}{2008}).

\bibitem[{\citenamefont{Huh and Scriven}(1971)}]{huh_scriven_1971}
\bibinfo{author}{\bibfnamefont{C.}~\bibnamefont{Huh}} \bibnamefont{and}
  \bibinfo{author}{\bibfnamefont{L.~E.} \bibnamefont{Scriven}},
  \bibinfo{journal}{J. Coll. Int. Sci.} \textbf{\bibinfo{volume}{35}},
  \bibinfo{pages}{85} (\bibinfo{year}{1971}).

\bibitem[{\citenamefont{Oron et~al.}(1997)\citenamefont{Oron, Davis, and
  Bankoff}}]{ODB97}
\bibinfo{author}{\bibfnamefont{A.}~\bibnamefont{Oron}},
  \bibinfo{author}{\bibfnamefont{S.~H.} \bibnamefont{Davis}}, \bibnamefont{and}
  \bibinfo{author}{\bibfnamefont{S.~G.} \bibnamefont{Bankoff}},
  \bibinfo{journal}{Rev. Mod. Phys.} \textbf{\bibinfo{volume}{69}},
  \bibinfo{pages}{931} (\bibinfo{year}{1997}).

\bibitem[{\citenamefont{Prevost et~al.}(1999)\citenamefont{Prevost, Rolley, and
  Guthmann}}]{Prevost99}
\bibinfo{author}{\bibfnamefont{A.}~\bibnamefont{Prevost}},
  \bibinfo{author}{\bibfnamefont{E.}~\bibnamefont{Rolley}}, \bibnamefont{and}
  \bibinfo{author}{\bibfnamefont{C.}~\bibnamefont{Guthmann}},
  \bibinfo{journal}{Phys. Rev. Lett.} \textbf{\bibinfo{volume}{83}},
  \bibinfo{pages}{348} (\bibinfo{year}{1999}).

\bibitem[{\citenamefont{Landau and Lifshitz}(1959)}]{landau_fluidmech_book}
\bibinfo{author}{\bibfnamefont{L.~D.} \bibnamefont{Landau}} \bibnamefont{and}
  \bibinfo{author}{\bibfnamefont{E.~M.} \bibnamefont{Lifshitz}},
  \emph{\bibinfo{title}{{Fluid Mechanics}}} (\bibinfo{publisher}{Pergamon},
  \bibinfo{year}{1959}).

\bibitem[{\citenamefont{Flekk{\o}y and
  Rothman}(1996)}]{flekkoy_fluctuating_1996}
\bibinfo{author}{\bibfnamefont{E.~G.} \bibnamefont{Flekk{\o}y}}
  \bibnamefont{and} \bibinfo{author}{\bibfnamefont{D.~H.}
  \bibnamefont{Rothman}}, \bibinfo{journal}{Phys. Rev. E}
  \textbf{\bibinfo{volume}{53}}, \bibinfo{pages}{1622} (\bibinfo{year}{1996}).

\bibitem[{\citenamefont{De~Zarate and Sengers}(2006)}]{zaratebook}
\bibinfo{author}{\bibfnamefont{J.~M.~O.} \bibnamefont{De~Zarate}}
  \bibnamefont{and} \bibinfo{author}{\bibfnamefont{J.~V.}
  \bibnamefont{Sengers}}, \emph{\bibinfo{title}{Hydrodynamic fluctuations in
  fluids and fluid mixtures}} (\bibinfo{publisher}{Elsevier},
  \bibinfo{year}{2006}).

\bibitem[{\citenamefont{Davidovitch et~al.}(2005)\citenamefont{Davidovitch,
  Moro, and Stone}}]{davidovitch_spreading_2005}
\bibinfo{author}{\bibfnamefont{B.}~\bibnamefont{Davidovitch}},
  \bibinfo{author}{\bibfnamefont{E.}~\bibnamefont{Moro}}, \bibnamefont{and}
  \bibinfo{author}{\bibfnamefont{H.~A.} \bibnamefont{Stone}},
  \bibinfo{journal}{Phys. Rev. Lett.} \textbf{\bibinfo{volume}{95}},
  \bibinfo{pages}{244505} (\bibinfo{year}{2005}).

\bibitem[{\citenamefont{Gr{\"u}n et~al.}(2006)\citenamefont{Gr{\"u}n, Mecke,
  and Rauscher}}]{gruen_thin-film_2006}
\bibinfo{author}{\bibfnamefont{G.}~\bibnamefont{Gr{\"u}n}},
  \bibinfo{author}{\bibfnamefont{K.}~\bibnamefont{Mecke}}, \bibnamefont{and}
  \bibinfo{author}{\bibfnamefont{M.}~\bibnamefont{Rauscher}},
  \bibinfo{journal}{J. Stat. Phys.} \textbf{\bibinfo{volume}{122}},
  \bibinfo{pages}{1261} (\bibinfo{year}{2006}).

\bibitem[{\citenamefont{Rauscher and Dietrich}(2008)}]{rauscher_wetting_2008}
\bibinfo{author}{\bibfnamefont{M.}~\bibnamefont{Rauscher}} \bibnamefont{and}
  \bibinfo{author}{\bibfnamefont{S.}~\bibnamefont{Dietrich}},
  \bibinfo{journal}{Annu. Rev. Mater. Res.} \textbf{\bibinfo{volume}{38}},
  \bibinfo{pages}{143} (\bibinfo{year}{2008}).

\bibitem[{\citenamefont{Belardinelli et~al.}(2016)\citenamefont{Belardinelli,
  Sbragaglia, Gross, and Andreotti}}]{belardinelli2016thermal}
\bibinfo{author}{\bibfnamefont{D.}~\bibnamefont{Belardinelli}},
  \bibinfo{author}{\bibfnamefont{M.}~\bibnamefont{Sbragaglia}},
  \bibinfo{author}{\bibfnamefont{M.}~\bibnamefont{Gross}}, \bibnamefont{and}
  \bibinfo{author}{\bibfnamefont{B.}~\bibnamefont{Andreotti}},
  \bibinfo{journal}{Physical Review E} \textbf{\bibinfo{volume}{94}},
  \bibinfo{pages}{052803} (\bibinfo{year}{2016}).

\bibitem[{\citenamefont{Nesic et~al.}(2015{\natexlab{a}})\citenamefont{Nesic,
  Cuerno, Moro, and Kondic}}]{nesic_dynamics_2015}
\bibinfo{author}{\bibfnamefont{S.}~\bibnamefont{Nesic}},
  \bibinfo{author}{\bibfnamefont{R.}~\bibnamefont{Cuerno}},
  \bibinfo{author}{\bibfnamefont{E.}~\bibnamefont{Moro}}, \bibnamefont{and}
  \bibinfo{author}{\bibfnamefont{L.}~\bibnamefont{Kondic}},
  \bibinfo{journal}{Eur. Phys. J. Special Topics}
  \textbf{\bibinfo{volume}{224}}, \bibinfo{pages}{379}
  (\bibinfo{year}{2015}{\natexlab{a}}).

\bibitem[{\citenamefont{Nesic et~al.}(2015{\natexlab{b}})\citenamefont{Nesic,
  Cuerno, Moro, and Kondic}}]{nesic_nonlinear_2015}
\bibinfo{author}{\bibfnamefont{S.}~\bibnamefont{Nesic}},
  \bibinfo{author}{\bibfnamefont{R.}~\bibnamefont{Cuerno}},
  \bibinfo{author}{\bibfnamefont{E.}~\bibnamefont{Moro}}, \bibnamefont{and}
  \bibinfo{author}{\bibfnamefont{L.}~\bibnamefont{Kondic}},
  \bibinfo{journal}{Phys. Rev. E} \textbf{\bibinfo{volume}{92}},
  \bibinfo{pages}{061002(R)} (\bibinfo{year}{2015}{\natexlab{b}}).

\bibitem[{\citenamefont{Fisher}(1984)}]{fisher_walks_1984}
\bibinfo{author}{\bibfnamefont{M.~E.} \bibnamefont{Fisher}},
  \bibinfo{journal}{J. Stat. Phys.} \textbf{\bibinfo{volume}{34}},
  \bibinfo{pages}{667} (\bibinfo{year}{1984}).

\bibitem[{\citenamefont{Bricmont et~al.}(1986)\citenamefont{Bricmont,
  El~Mellouki, and Fr{\"o}hlich}}]{bricmont_random_1986}
\bibinfo{author}{\bibfnamefont{J.}~\bibnamefont{Bricmont}},
  \bibinfo{author}{\bibfnamefont{A.}~\bibnamefont{El~Mellouki}},
  \bibnamefont{and}
  \bibinfo{author}{\bibfnamefont{J.}~\bibnamefont{Fr{\"o}hlich}},
  \bibinfo{journal}{J. Stat. Phys.} \textbf{\bibinfo{volume}{42}},
  \bibinfo{pages}{743} (\bibinfo{year}{1986}).

\bibitem[{\citenamefont{Lebowitz and Maes}(1987)}]{lebowitz_effect_1987}
\bibinfo{author}{\bibfnamefont{J.~L.} \bibnamefont{Lebowitz}} \bibnamefont{and}
  \bibinfo{author}{\bibfnamefont{C.}~\bibnamefont{Maes}}, \bibinfo{journal}{J.
  Stat. Phys.} \textbf{\bibinfo{volume}{46}}, \bibinfo{pages}{39}
  (\bibinfo{year}{1987}).

\bibitem[{\citenamefont{Crassous and Charlaix}(1994)}]{Crassous}
\bibinfo{author}{\bibfnamefont{J.}~\bibnamefont{Crassous}} \bibnamefont{and}
  \bibinfo{author}{\bibfnamefont{E.}~\bibnamefont{Charlaix}},
  \bibinfo{journal}{Europhys. Lett.} \textbf{\bibinfo{volume}{28}},
  \bibinfo{pages}{415} (\bibinfo{year}{1994}).

\bibitem[{\citenamefont{Joanny and de~Gennes}(1984)}]{JdG84}
\bibinfo{author}{\bibfnamefont{J.-F.} \bibnamefont{Joanny}} \bibnamefont{and}
  \bibinfo{author}{\bibfnamefont{P.-G.} \bibnamefont{de~Gennes}},
  \bibinfo{journal}{J. Chem. Phys.} \textbf{\bibinfo{volume}{81}},
  \bibinfo{pages}{552} (\bibinfo{year}{1984}).

\bibitem[{\citenamefont{de~Gennes}(1985)}]{deGennesRMP}
\bibinfo{author}{\bibfnamefont{P.~G.} \bibnamefont{de~Gennes}},
  \bibinfo{journal}{Rev. Mod. Phys.} \textbf{\bibinfo{volume}{57}},
  \bibinfo{pages}{827} (\bibinfo{year}{1985}).

\bibitem[{\citenamefont{de~Gennes et~al.}(2003)\citenamefont{de~Gennes,
  Brochard-Wyart, and Quere}}]{deGennes_book}
\bibinfo{author}{\bibfnamefont{P.-G.} \bibnamefont{de~Gennes}},
  \bibinfo{author}{\bibfnamefont{F.}~\bibnamefont{Brochard-Wyart}},
  \bibnamefont{and} \bibinfo{author}{\bibfnamefont{D.}~\bibnamefont{Quere}},
  \emph{\bibinfo{title}{Capillarity and Wetting Phenomena: Drops, Bubbles,
  Pearls, Waves}} (\bibinfo{publisher}{Springer, New York},
  \bibinfo{year}{2003}).

\bibitem[{\citenamefont{Perrin et~al.}(2016)\citenamefont{Perrin, Lhermerout,
  Davitt, Rolley, and Andreotti}}]{perrin_2016}
\bibinfo{author}{\bibfnamefont{H.}~\bibnamefont{Perrin}},
  \bibinfo{author}{\bibfnamefont{R.}~\bibnamefont{Lhermerout}},
  \bibinfo{author}{\bibfnamefont{K.}~\bibnamefont{Davitt}},
  \bibinfo{author}{\bibfnamefont{E.}~\bibnamefont{Rolley}}, \bibnamefont{and}
  \bibinfo{author}{\bibfnamefont{B.}~\bibnamefont{Andreotti}},
  \bibinfo{journal}{Phys. Rev. Lett.} \textbf{\bibinfo{volume}{116}},
  \bibinfo{pages}{184502} (\bibinfo{year}{2016}).

\bibitem[{\citenamefont{Larson}(1999)}]{Larson}
\bibinfo{author}{\bibfnamefont{R.~G.} \bibnamefont{Larson}},
  \emph{\bibinfo{title}{The structure and rheology of complex fluids}}
  (\bibinfo{publisher}{Oxford University Press}, \bibinfo{year}{1999}).

\bibitem[{\citenamefont{Hanggi et~al.}(1990)\citenamefont{Hanggi, Talkner, and
  Borkovec}}]{HanggiReview}
\bibinfo{author}{\bibfnamefont{P.}~\bibnamefont{Hanggi}},
  \bibinfo{author}{\bibfnamefont{P.}~\bibnamefont{Talkner}}, \bibnamefont{and}
  \bibinfo{author}{\bibfnamefont{M.}~\bibnamefont{Borkovec}},
  \bibinfo{journal}{Rev. Mod. Phys.} \textbf{\bibinfo{volume}{62}},
  \bibinfo{pages}{251} (\bibinfo{year}{1990}).

\bibitem[{\citenamefont{Toll}(1956)}]{Toll}
\bibinfo{author}{\bibfnamefont{J.~S.} \bibnamefont{Toll}},
  \bibinfo{journal}{Phys. Rev.} \textbf{\bibinfo{volume}{104}},
  \bibinfo{pages}{1760} (\bibinfo{year}{1956}).

\bibitem[{\citenamefont{Zwanzig}(2001)}]{Zwanzig}
\bibinfo{author}{\bibfnamefont{R.}~\bibnamefont{Zwanzig}},
  \emph{\bibinfo{title}{Nonequilibrium statistical mechanics}}
  (\bibinfo{publisher}{Oxford University Press}, \bibinfo{year}{2001}).

\bibitem[{\citenamefont{Safran}(2003)}]{Saf03}
\bibinfo{author}{\bibfnamefont{S.~A.} \bibnamefont{Safran}},
  \emph{\bibinfo{title}{Statistical thermodynamics of surfaces, interfaces, and
  membranes}} (\bibinfo{publisher}{Westview Press, Boulder, CO},
  \bibinfo{year}{2003}).

\bibitem[{\citenamefont{Voinov}(1976)}]{COXVOINOV}
\bibinfo{author}{\bibfnamefont{O.~V.} \bibnamefont{Voinov}},
  \bibinfo{journal}{Fluid Dyn.} \textbf{\bibinfo{volume}{11}},
  \bibinfo{pages}{714} (\bibinfo{year}{1976}).

\bibitem[{\citenamefont{Snoeijer et~al.}(2007)\citenamefont{Snoeijer,
  Andreotti, Delon, and Fermigier}}]{snoeijer_2007b}
\bibinfo{author}{\bibfnamefont{J.~H.} \bibnamefont{Snoeijer}},
  \bibinfo{author}{\bibfnamefont{B.}~\bibnamefont{Andreotti}},
  \bibinfo{author}{\bibfnamefont{G.}~\bibnamefont{Delon}}, \bibnamefont{and}
  \bibinfo{author}{\bibfnamefont{m.}~\bibnamefont{Fermigier}},
  \bibinfo{journal}{J. Fluid. Mech.} \textbf{\bibinfo{volume}{579}},
  \bibinfo{pages}{63} (\bibinfo{year}{2007}).

\bibitem[{\citenamefont{Gradshteyn and Ryzhik}(2000)}]{GR}
\bibinfo{author}{\bibfnamefont{S.}~\bibnamefont{Gradshteyn}} \bibnamefont{and}
  \bibinfo{author}{\bibfnamefont{I.~M.} \bibnamefont{Ryzhik}},
  \emph{\bibinfo{title}{Table of Integrals, Series, and Products}}
  (\bibinfo{publisher}{Academic, San Diego}, \bibinfo{year}{2000}).

\bibitem[{\citenamefont{Benzi et~al.}(1981)\citenamefont{Benzi, Sutera, and
  Vulpiani}}]{Stochastic1}
\bibinfo{author}{\bibfnamefont{R.}~\bibnamefont{Benzi}},
  \bibinfo{author}{\bibfnamefont{A.}~\bibnamefont{Sutera}}, \bibnamefont{and}
  \bibinfo{author}{\bibfnamefont{A.}~\bibnamefont{Vulpiani}},
  \bibinfo{journal}{J. Phys. A: Math. Gen.} \textbf{\bibinfo{volume}{14}},
  \bibinfo{pages}{L453} (\bibinfo{year}{1981}).

\bibitem[{\citenamefont{Gammaitoni et~al.}(1998)\citenamefont{Gammaitoni,
  Hanggi, Jung, and Marchesoni}}]{Stochastic2}
\bibinfo{author}{\bibfnamefont{L.}~\bibnamefont{Gammaitoni}},
  \bibinfo{author}{\bibfnamefont{P.}~\bibnamefont{Hanggi}},
  \bibinfo{author}{\bibfnamefont{P.}~\bibnamefont{Jung}}, \bibnamefont{and}
  \bibinfo{author}{\bibfnamefont{F.}~\bibnamefont{Marchesoni}},
  \bibinfo{journal}{Rev. Mod. Phys.} \textbf{\bibinfo{volume}{70}},
  \bibinfo{pages}{223} (\bibinfo{year}{1998}).

\bibitem[{\citenamefont{Martens et~al.}(2012)\citenamefont{Martens, Bocquet,
  and Barrat}}]{Martens}
\bibinfo{author}{\bibfnamefont{K.}~\bibnamefont{Martens}},
  \bibinfo{author}{\bibfnamefont{L.}~\bibnamefont{Bocquet}}, \bibnamefont{and}
  \bibinfo{author}{\bibfnamefont{J.-L.} \bibnamefont{Barrat}},
  \bibinfo{journal}{Soft Matter} \textbf{\bibinfo{volume}{8}},
  \bibinfo{pages}{4197} (\bibinfo{year}{2012}).

\bibitem[{\citenamefont{Mansard et~al.}(2011)\citenamefont{Mansard, Colin,
  Chauduri, and Bocquet}}]{Mansard}
\bibinfo{author}{\bibfnamefont{V.}~\bibnamefont{Mansard}},
  \bibinfo{author}{\bibfnamefont{A.}~\bibnamefont{Colin}},
  \bibinfo{author}{\bibfnamefont{P.}~\bibnamefont{Chauduri}}, \bibnamefont{and}
  \bibinfo{author}{\bibfnamefont{L.}~\bibnamefont{Bocquet}},
  \bibinfo{journal}{Soft Matter} \textbf{\bibinfo{volume}{7}},
  \bibinfo{pages}{5524} (\bibinfo{year}{2011}).

\bibitem[{\citenamefont{Picard et~al.}(2002)\citenamefont{Picard, Ajdari,
  Bocquet, and Lequeux}}]{Picard}
\bibinfo{author}{\bibfnamefont{G.}~\bibnamefont{Picard}},
  \bibinfo{author}{\bibfnamefont{A.}~\bibnamefont{Ajdari}},
  \bibinfo{author}{\bibfnamefont{L.}~\bibnamefont{Bocquet}}, \bibnamefont{and}
  \bibinfo{author}{\bibfnamefont{F.}~\bibnamefont{Lequeux}},
  \bibinfo{journal}{Phys. rev. E} \textbf{\bibinfo{volume}{66}},
  \bibinfo{pages}{051501} (\bibinfo{year}{2002}).

\bibitem[{\citenamefont{Hebraud and Lequeux}(1997)}]{HL98}
\bibinfo{author}{\bibfnamefont{P.}~\bibnamefont{Hebraud}} \bibnamefont{and}
  \bibinfo{author}{\bibfnamefont{F.}~\bibnamefont{Lequeux}},
  \bibinfo{journal}{Phys. Rev. Lett.} \textbf{\bibinfo{volume}{81}},
  \bibinfo{pages}{293} (\bibinfo{year}{1997}).

\bibitem[{\citenamefont{Sollich et~al.}(1997)\citenamefont{Sollich, Lequeux,
  Hebraud, and Cates}}]{Sollich1}
\bibinfo{author}{\bibfnamefont{P.}~\bibnamefont{Sollich}},
  \bibinfo{author}{\bibfnamefont{F.}~\bibnamefont{Lequeux}},
  \bibinfo{author}{\bibfnamefont{P.}~\bibnamefont{Hebraud}}, \bibnamefont{and}
  \bibinfo{author}{\bibfnamefont{M.~E.} \bibnamefont{Cates}},
  \bibinfo{journal}{Phys. Rev. Lett.} \textbf{\bibinfo{volume}{78}},
  \bibinfo{pages}{2020} (\bibinfo{year}{1997}).

\bibitem[{\citenamefont{Sollich}(1998)}]{Sollich2}
\bibinfo{author}{\bibfnamefont{P.}~\bibnamefont{Sollich}},
  \bibinfo{journal}{Phys. Rev. E} \textbf{\bibinfo{volume}{58}},
  \bibinfo{pages}{738} (\bibinfo{year}{1998}).

\bibitem[{\citenamefont{Sollich and Cates}(2012)}]{Sollich3}
\bibinfo{author}{\bibfnamefont{P.}~\bibnamefont{Sollich}} \bibnamefont{and}
  \bibinfo{author}{\bibfnamefont{M.~E.} \bibnamefont{Cates}},
  \bibinfo{journal}{Phys. Rev. E} \textbf{\bibinfo{volume}{85}},
  \bibinfo{pages}{031127} (\bibinfo{year}{2012}).

\bibitem[{\citenamefont{Bonn et~al.}(2017)\citenamefont{Bonn, Denn, Berthier,
  Divoux, and Manneville}}]{BonnReviewYIELDSTRESS}
\bibinfo{author}{\bibfnamefont{D.}~\bibnamefont{Bonn}},
  \bibinfo{author}{\bibfnamefont{M.~M.} \bibnamefont{Denn}},
  \bibinfo{author}{\bibfnamefont{L.}~\bibnamefont{Berthier}},
  \bibinfo{author}{\bibfnamefont{T.}~\bibnamefont{Divoux}}, \bibnamefont{and}
  \bibinfo{author}{\bibfnamefont{S.}~\bibnamefont{Manneville}},
  \bibinfo{journal}{arXiv:1502.05281.v2} pp. \bibinfo{pages}{1--38}
  (\bibinfo{year}{2017}).

\end{thebibliography}
\end{document}